%% file: paper.tex
\documentclass[a4paper,fleqn,usenatbib,useAMS]{mn2e} 
\usepackage{graphicx} 
\usepackage{subfigure} 
\usepackage{amsmath} 
\usepackage{amssymb} 
\usepackage{amsfonts}
\usepackage{comment} 
\usepackage{color} 
\usepackage[dvipsnames]{xcolor} 
 
\usepackage{times} 
\usepackage{natbib} 
 
\usepackage{soul} 
\usepackage{ulem} 
 
\newif\ifAMStwofonts 
\AMStwofontstrue 

\newcommand{\ud}{\mathrm{d}} 
\newcommand{\p}{\partial} 
\newcommand{\cH}{\mathcal{H}}

\newcommand{\mathmat}[1]{\boldsymbol{\mathsf{#1}}}
 
\def\ltsima{$\; \buildrel < \over \sim \;$} 
\def\simlt{\lower.5ex\hbox{\ltsima}} 
\def\gtsima{$\; \buildrel > \over \sim \;$} 
\def\simgt{\lower.5ex\hbox{\gtsima}} 
 
\title[The LIGER method]{LIGER: mock relativistic light-cones from Newtonian simulations} 
\author[Borzyszkowski, Bertacca \& Porciani]{Mikolaj Borzyszkowski\thanks{E-mail: mikolajb@uni-bonn.de}, Daniele Bertacca and Cristiano Porciani\\ 
Argelander-Institut f\"ur Astronomie, Auf dem H\"ugel 71, D-53121 Bonn, Germany} 
\begin{document} 
 
\include{./adsmacro}

\date{\today} 
 
\pagerange{\pageref{firstpage}--\pageref{lastpage}} \pubyear{2015} 
 
\maketitle 
 
\label{firstpage} 
 
\begin{abstract} 
We introduce a method to create mock galaxy catalogues in redshift space including general relativistic effects to linear order in the cosmological perturbations.  
We dub our method LIGER, short for `light cones with general relativity'.   
LIGER takes a (N-body or hydrodynamic) Newtonian simulation as an input and outputs the distribution of galaxies in comoving redshift space. 
This result is achieved making use of a coordinate transformation and simultaneously accounting for lensing magnification.  
The calculation includes both local corrections and terms that have been integrated along the line of sight.  
Our fast implementation allows the production of many realizations that can be used to forecast the performance of forthcoming wide-angle surveys and to estimate the covariance matrix of the observables.  
To facilitate this use, we also present a variant of LIGER designed for large-volume simulations with low mass resolution.  
In this case, the galaxy distribution on large scales is obtained by biasing the matter-density field.  
Finally, we present two sample applications of LIGER.  
First, we discuss the impact of weak gravitational lensing onto the angular clustering of galaxies in a Euclid-like survey. 
In agreement with previous analytical studies, we find that magnification bias can be measured with high confidence.
Second, we focus on two generally neglected Doppler-induced effects: magnification and the change of number counts with redshift.
We show that the corresponding redshift-space distortions can be detected at 5.5$\sigma$ significance with the completed Square Kilometre Array. 
\end{abstract} 
 
 
\section{Introduction} 
The advent of galaxy redshift surveys has revolutionised our understanding of the large-scale structure of the Universe and provided us with multiple ways to constrain the cosmological model.  
Mock catalogues of synthetic galaxies play a threefold role in the analysis of these datasets \citep{Cole:1998,Blaizot:2005,Kitzbichler:2007,Sousbie:2008,Carlson:2010,Merson:2013}. (i) They shape 
theoretical predictions into structures that closely match observations. (ii) They form a straightforward tool 
to derive biases and covariance matrices of estimators for statistical descriptions of the large-scale structure 
(e.g. correlation functions or their Fourier analogues). 
(iii) Related to that, as forecasting tools, they provide key information to designing new surveys by 
 minimising the impact of statistical errors and systematic effects on selected observables.  
 
Since the 1970s, the size of galaxy catalogues has constantly increased in terms of solid-angle and redshift coverage as well as in sampling rate.  
The next generation of surveys will provide us with the possibility to measure galaxy clustering on scales comparable with the Hubble radius \citep[e.g.][]{EuclidRedBook, Levi:2013Desi, Maartens:2015SKA}.  Theoretical studies suggest that a number of general relativistic effects might be detectable on these scales. In order to fully exploit the potential of the new datasets, 
it is therefore imperative to develop analysis tools (and thus mock catalogues) that include these effects. 
In this paper, we present a method to create mock galaxy catalogues that incorporate relativistic corrections 
and are built upon the output of either common Newtonian simulations of galaxy formation or semi-analytic models based on standard N-body simulations. 
 
Relativistic effects arise from the fact that we observe galaxies on our past lightcone. 
The presence of perturbations superimposed to a Friedmann-Robertson-Walker (FRW) background alters the 
null geodesics of the photons emitted by distant galaxies. In consequence, all the direct observables for a galaxy   
are different than in a smooth universe: its redshift, angular position on the sky and the flux in any given waveband. 
Galaxy peculiar velocities, for instance, distort the radial pattern of the galaxy distribution 
 \citep{Kaiser:1987qv, Hamilton:1997zq}.  
Similarly, magnification due to gravitational lensing  
modifies the observed number counts in flux-limited samples \citep{Turner1980, Turner:1984ch, Sasaki:1987ad, Matsubara:2000pr}.  
Many recent studies have demonstrated the existence of several additional corrections that, although suppressed on smaller scales, might generate observable signals on distances comparable with the Hubble radius \citep[e.g.][]{McDonald2009, Yoo:2008tj, Yoo:2009au, Bonvin:2011bg, Challinor:2011bk, Bertacca:2012tp, Jeong:2011as, Yoo:2012se, DiDio:2013bqa, DiDio:2013sea, Bonvinetal2014, Montanari:2015rga, Yoo:2013zga, Bonvin:2015kuc, Cardona:2016qxn, DiDio:2016ykq, Raccanelli:2016avd, Raccanelli:2013gja, Raccanelli:2015vla, Gaztanaga:2015jrs}.  
At linear order in the perturbations, these additional corrections include Doppler terms plus Sachs-Wolfe \citep[standard and integrated, see][]{Sachs:1967er, Rees:1968} and (Shapiro) time-delay contributions. 
Robust models of galaxy clustering on large scales should thus include these modifications that, most likely, 
will be key to extracting unbiased  
information on the dark sector of the Universe (i.e. on the nature of dark energy and dark matter) and to improve constraints on primordial non-Gaussianity. 
This can be done following different approaches. One possibility is to study structure formation 
using N-body simulations that include dynamical space-time  
variables in the weak-field approximation  \citep{Adamek:2013wja, Adamek:2014xba, Adamek:2015eda, Adamek:2016zes}, within a post-Friedmann framework  \citep{Milillo:2015cva, Bruni:2013mua, Thomas:2015kua}, or by employing full numerical relativity \citep{Bentivegna:2016, Giblin-Mertens-Starkman-2016}. 
Alternatively, one can correct a posteriori the results of Newtonian simulations to account for lightcone effects \citep{Chisari:2011iq}. This is feasible because, at linear order in the perturbations, the mathematical description of a pressureless fluid can be formulated so that 
there is agreement between general relativity and its Newtonian approximation \citep{Haugg:2012ng, Rigopoulos:2013nda, Fidler:2015npa, Fidler:2016tir} 
 
\begin{figure} 
  \centering 
  \includegraphics[width=0.48\textwidth,bb=0 0 694 369,keepaspectratio=true]{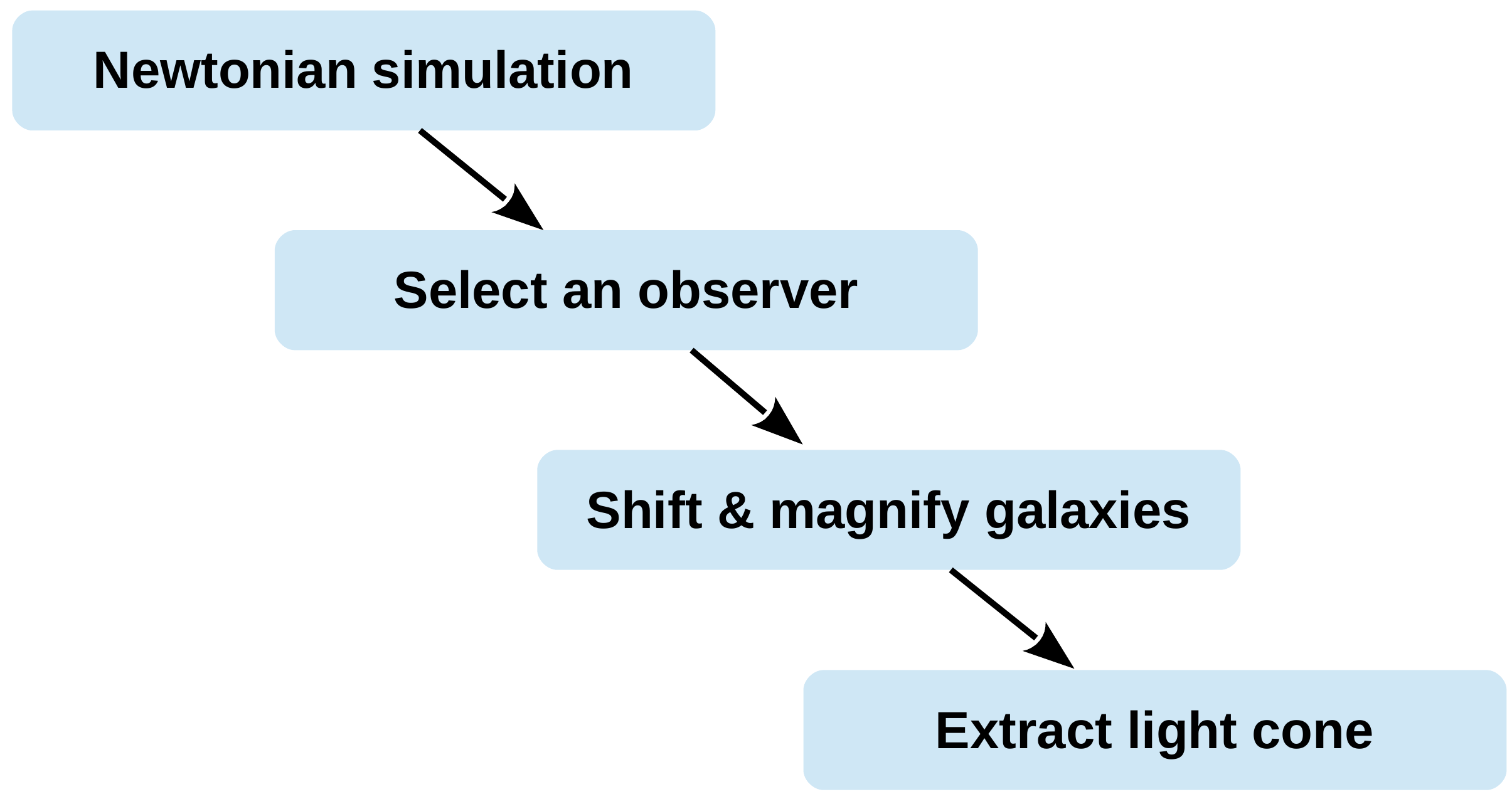}
  \caption{Schematic diagram illustrating the sequence of main processing steps in the LIGER method. } 
  \label{pic_diagram} 
\end{figure}

In this paper, we follow the latter approach to develop the LIGER (light cones using general relativity) method. 
As illustrated in Fig. \ref{pic_diagram}, schematically, LIGER takes a Newtonian simulation as an input and, 
after selecting an observer, outputs the distribution of galaxies in `comoving redshift space' (i.e. as it would be inferred by applying the background metric to the observed galaxy properties). 
The algorithm combines 
the original snapshots of the simulation at constant background time to produce the galaxy 
distribution on the perturbed light cone.    
This is achieved by using a coordinate transformation that includes local terms and contributions that are integrated along the line of sight. 
 
Multiple efforts have been made in the literature to investigate the detectability of subtle relativistic effects from forthcoming survey data. 
Generally, these studies are based on the Fisher-information matrix, use idealised survey characteristics and neglect systematics. 
The ultimate test to discern which relativistic effects will be observable is to apply the very same estimators that are used for the data to the LIGER mocks. 
This exciting perspective provides the main motivation for our work. 
The numerical implementation of the LIGER method will be made publicly available in due course.  
 
The paper is structured as follows. In Section 2, we introduce the LIGER method and describe its numerical implementation. In Section 3, we present two straightforward applications of our code. As an illustration of LIGER's functionality, we first re-analyse a result which has already been discussed in the literature, namely, the impact of magnification bias in the observed cross-correlation of galaxy samples at substantially different redshifts. Subsequently, we discuss the more challenging detection of Doppler terms in the galaxy angular power spectrum at low redshift.  
Finally, in Section 4, we conclude. 
Throughout, we adopt units in which the speed of light is one and define the space-time metric tensor to have  
signature $(-,+,+,+)$. Greek indices indicate space-time components (i.e. run from 0 to 3) while Latin indices label spatial components (i.e. run 
from 1 to 3). The Einstein summation convention is adopted.

\section{The LIGER method} \subsection{Theory} 
\subsubsection{Redshift-space distortions} 
We observe galaxies as they are at the time in which their worldline intersects our past light cone.
The comoving location of a galaxy can be inferred from two basic observables: its position on the sky, ${\bf n}_{\rm s}$ (a unit vector defined in terms of two angles), and its redshift, $z$.  
In fact, these data are sufficient to build three-dimensional maps of the galaxy distribution provided that we assume to live in an unperturbed FRW universe with a fixed set of cosmological parameters. 
In reality, such a `redshift-space' map gives a distorted portrayal of the cosmic web due to the presence of inhomogeneities \citep{Sargent-Turner-1977}. 
Galaxies are artificially shifted both in the radial and tangential directions due to their peculiar motions and the bending of the light they emit. 
These effects are collectively known under the name of redshift-space distortions. 
  
In mathematical terms, redshift-space is characterized by a set of coordinates that `flatten' our past light cone \citep[e.g.][]{Bertacca:2015}. 
For instance, the null geodesic from an observed galaxy to us can be described in terms of the following conformal space-time coordinates: 
\begin{equation} 
x_{\rm s}^\mu=(\eta_{\rm s},\;  {\bf x}_{\rm s})=(\eta_0- \chi_{\rm s}, \;  \chi_{\rm s} \, {\bf n}_{\rm s})\;, 
\end{equation} 
where  $\eta_0$ is the present-day value of conformal time (i.e. at observation), $\chi_{\rm s}$ denotes the comoving distance (from the observer) of events located along the geodesic in the unperturbed model universe and $n^i_{\rm s}=x_{\rm s}^i/ \chi_{\rm s}$. 
The full distance to the galaxy corresponds to the observed redshift $z$, in compact notation $\chi_{\rm s}(z)$.

\begin{figure} 
\centering 
\includegraphics[width=16 cm,bb=100 0 1200 580]{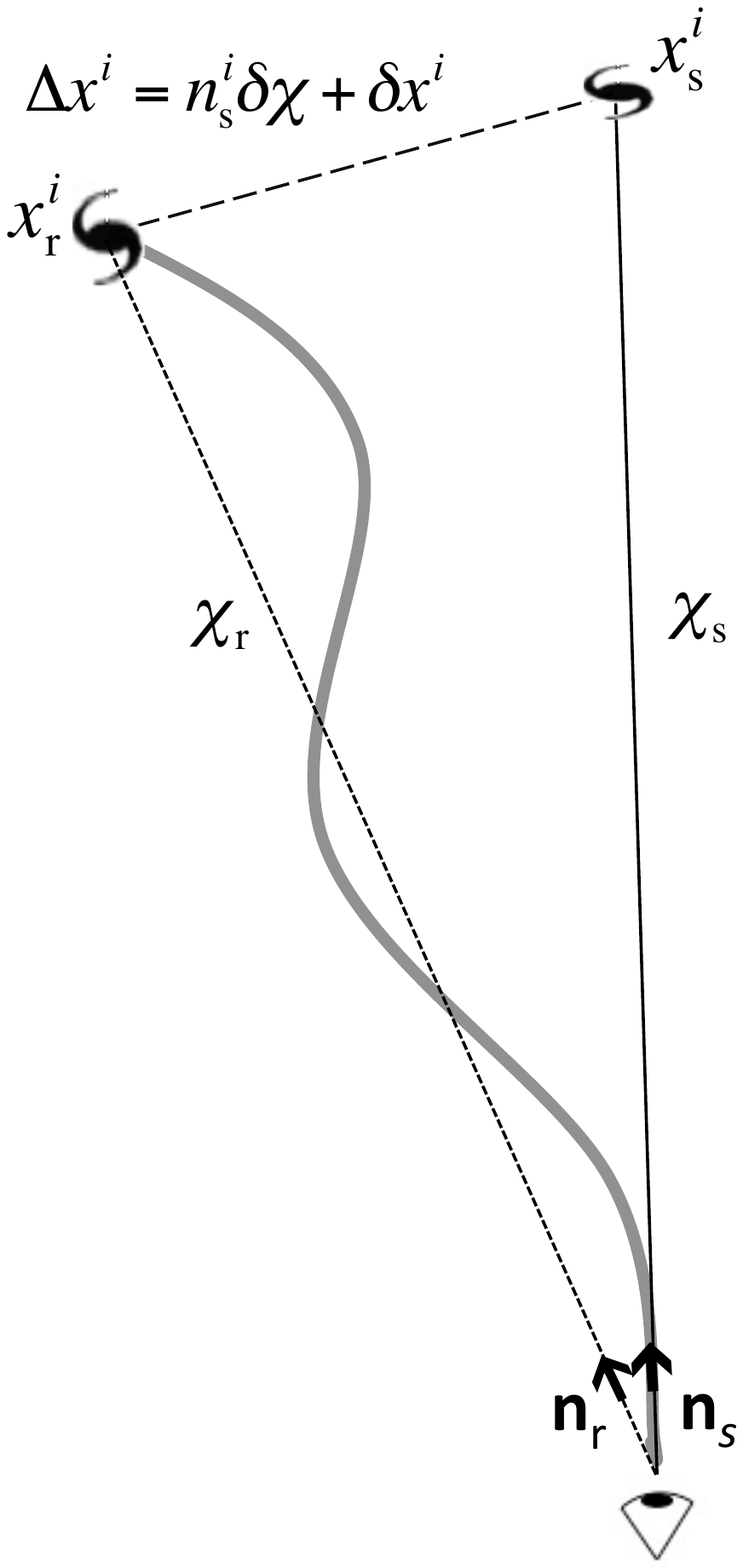} 
\caption{ 
Real- and redshift-space perspectives.  
A galaxy with real-space position $x^i_{\rm r}$ (top left), located at distance $\chi_{\rm r}$ from the observer (bottom), 
is assigned an apparent position $x^i_{\rm s}$ in redshift space (top right) at distance $\chi_{\rm s}$. 
Since the photon path to the observer in real space is not straight, the observed position of the galaxy on the sky, ${\bf n}_{\rm s}$ does 
not coincide with its actual one, ${\bf n}_{\rm r}$. 
\label{fig:1}} 
\end{figure}

For a given photon path (see Fig.~\ref{fig:1}), we want to define a mapping from real to redshift space, 
\begin{equation} 
x_{\rm r}^\mu [\chi_{\rm r}(\chi_{\rm s})] = x_{\rm s}^\mu (\chi_{\rm s})+ \Delta x^\mu (\chi_{\rm s})\;, 
\end{equation} 
where 
$x_{\rm r}^i$ denotes the actual comoving position located at distance $\chi_{\rm r}$  along the direction  $n^i_{\rm r}=x_{\rm r}^i/\chi_{\rm r}$. 
[In general, we use the subscripts `s' and `r' to distinguish redshift-space quantities from their real-space counterparts.]  
Perturbing $x^\mu_{\rm r}$ around $x^\mu_{\rm s}$ and writing $\chi_{\rm r} = \chi_{\rm s}+ \delta \chi$, we obtain, at linear order, 
 \begin{eqnarray} 
x_{\rm r}^\mu ( \chi_{\rm r}) &=& x_{\rm s}^\mu (\chi_{\rm r})+ \delta x^\mu  (\chi_{\rm r}) \nonumber \\ 
                            &=& x_{\rm s}^\mu (\chi_{\rm s})+ \frac{\ud  x_{\rm s}^\mu}{\ud\chi_{\rm s}} \delta \chi+\delta x^\mu (\chi_{\rm s})\;.
\end{eqnarray} 
By using $\chi_{\rm s}$ as the affine parameter for the null geodesic, we write the total derivative along the past light cone as $\ud /\ud \chi_{\rm s}= - \p / \p \eta_{\rm s} + n^i_{\rm s} \p/\p  x_{\rm s}^i$.
Since ${\ud  x_{\rm s}^i}/{\ud\chi_{\rm s}}=n^i_{\rm s}$ [to zero order], linear redshift-space distortions can be written as  
\begin{eqnarray} 
\label{Deltat} 
\Delta x^0(\chi_{\rm s})&=&-\delta\chi+\delta x^0(\chi_{\rm s}) \\
\label{Deltax} 
\Delta x^i ( \chi_{\rm s}) &=& n^i_{\rm s} \delta \chi+ \delta x^i ( \chi_{\rm s}) \;. 
\end{eqnarray} 
The first term on the right-hand side of equation~(\ref{Deltax}) corresponds to the change in the affine parameter while the second one derives from the  
perturbation of the photon path and has both tangential and radial components.
Note that the real-space distance to the galaxy does not coincide with $\chi_{\rm s}(z)$.

\subsubsection{Perturbations of the past light cone} 
In order to compute explicitly all the terms in equations~(\ref{Deltat}) and (\ref{Deltax}), we need to work out how metric perturbations alter null geodesics.  
In what follows, we model the matter content of the universe as a collisionless fluid in the single-stream regime. This common assumption provides a suitable 
approximation on the large scales we are interested in. We use the subscripts `e'  and `o' to denote the fluid properties evaluated at the position of the light source (the galaxy) when the photons were emitted and at the location of the observer when the photons were received, respectively. 
For linear scalar perturbations in the Poisson gauge\footnote{The restricted Poisson gauge containing only scalar perturbations is also known as the longitudinal or conformal Newtonian gauge.}, the space-time metric can be expressed as  
\begin{equation}\label{New} 
\ud s^2=a^2(\eta)\Big[-\big(1+2 \Psi\big)\ud \eta^2+\big(1-2 \Phi \big)\delta_{ij} \ud x_{\rm r}^i \ud x_{\rm r}^j\Big]\;, 
\end{equation} 
where $a$ denotes the scale factor of the expanding universe while $\Psi$ and $\Phi$ indicate the Bardeen potentials of the inhomogeneities. 
In this framework,  the redshift of a galaxy measured by an observer is 
\begin{equation} 
 1+z= \frac{(u_\mu p^\mu)|_{\rm e}}{(u_\mu p^\mu)|_{\rm o}}\;, 
\end{equation} 
where $u^\mu$ denotes the four-velocity of the matter fluid (we assume there is no velocity bias) and $p^\mu$ is the photon four-momentum. 
By perturbing the photon geodesic around the FRW solution, we derive expressions for  
$\delta x^\mu$ and $\delta \chi$. 
The final result for the galaxy shift is \citep[see also][]{Yoo:2008tj, Yoo:2009au, Bonvin:2011bg, Challinor:2011bk, Jeong:2011as} 
 \begin{eqnarray} \label{eq_delta_los} 
\delta \chi \!\!\!\!\!&=&\!\!\!\!\! - \left(\chi_{\rm s}+{1 \over \cH}\right) \left[\Psi_{\rm o}-\left(n_{\rm s}^i v_i\right)_{\rm o} \right]+{1 \over \cH} \left[\Psi_{\rm e}-\left(n_{\rm s}^i v_i\right)_{\rm e}\right]   \nonumber \\ 
&&\!\!\!\!\!+\int_0^{\chi_{\rm s}}  \left[2 \Psi +(\chi_{\rm s}- \chi) \p_0 \left(\Phi+  \Psi \right) \right]\nonumber \,\ud \chi\\ 
&&\!\!\!\!\!+{1 \over \cH}\int_0^{\chi_{\rm s}}  \p_0 \left(\Phi+  \Psi \right)\,\ud \chi  \;,
\end{eqnarray} 
\begin{eqnarray} \label{eq_delta_len} 
\delta x^0\!\!\!\!\!&=&\!\!\!\!\!-\chi_{\rm s}\left[ \Psi_{\rm o}- \left(n_{\rm s}^i v_i\right)_{\rm o} \right]+ 2 \int_0^{\chi_{\rm s}}   \Psi \,\ud \chi \nonumber \\
&&\!\!\!\!\!+  \int_0^{\chi_{\rm s}} \left(\chi_{\rm s} - \chi\right)  \p_0 \left(\Phi+  \Psi \right)\,\ud \chi \;,\\
\delta x^i\!\!\!\!\!&=&\!\!\!\!\!-\left( v^i_{\rm o} + \Phi_{\rm o} n^i_{\rm s}\right) \chi_{\rm s} +2 n^i_{\rm s} \int_0^{\chi_{\rm s}}  \Phi \,\ud \chi\nonumber \\ 
&&\!\!\!\!\! - \int_0^{\chi_{\rm s}}  (\chi_{\rm s}-\chi)  \delta^{ij} \p_j \left( \Phi + \Psi \right)\,\ud \chi \;, 
\end{eqnarray} 
where $\cH= \p_0 \ln a$ and $\chi_{\rm s}$ are evaluated at the observed redshift of the galaxy while $v^i$ is the peculiar velocity.
Here, local corrections express the Sachs-Wolfe and the Doppler effects. Those integrated along the line of sight derive from
gravitational lensing, the Shapiro time-delay and the integrated Sachs-Wolfe effect.
 
\subsubsection{Magnification} 
Metric perturbations also alter the solid angle under which galaxies are seen by distant observers thereby enhancing or decreasing their apparent flux. 
In terms of the luminosity distance, $D_{\rm L}$, the magnification of a galaxy is defined as
\begin{equation} 
 {\mathcal M}=\left( \frac{D_{\rm L}}{\bar{D}_{\rm L}}\right)^{-2}\;, 
\end{equation} 
where $\bar{D}_{\rm L}$ denotes the luminosity distance in the background model universe evaluated at
the observed redshift of the galaxy. 
 
At linear order in the perturbations, we can write 
\citep[e.g.][]{Challinor:2011bk,Bertacca:2015} 
\begin{eqnarray} 
\label{eq:magnif} 
 {\mathcal M}\!\!\!\!\!&=&\!\!\!\!\!1-2\left(1-\frac{1}{{\mathcal H}\chi_{\rm s}}\right)\left[\Psi_{\rm o}-\left(n_{\rm s}^i v_i\right)_{\rm o}\right] \nonumber\\ 
 &\,&\!\!\!\!\!+2\left(1-\frac{1}{{\mathcal H}\chi_{\rm s}}\right)\left[\int_0^{\chi_{\rm s}} \p_0(\Phi+\Psi) \,\ud \chi+\Psi_{\rm e}-\left(n^i_{\rm s}v_i\right)_{\rm e} \right]\nonumber\\ 
 &\,&\!\!\!\!\!+2\Phi_{\rm e}-\frac{2}{\chi_{\rm s}}\int_0^{\chi_{\rm s}} (\Phi+\Psi)\,\ud \chi+2\kappa\;,\label{eq_magni} 
\end{eqnarray} %
where
\begin{eqnarray}
 \kappa=\frac{1}{2}\int_0^{\chi_s} \left(\chi_{\rm s}- \chi \right) \frac{\chi}{\chi_{\rm s}}\nabla_\perp^2 \left(\Phi+\Psi\right) \ud \chi
\end{eqnarray}
corresponds to the classical convergence and the differential operator $\nabla_\perp^2$ is defined as 
\begin{eqnarray} 
 \nabla_\perp^2 =\nabla^2-\left(n^i_{\rm s} \p_i \right)^2-\frac{2}{\chi}n^i_{\rm s} \p_i\;. 
\end{eqnarray} 
Note that the magnification includes contributions from different physical effects. 
We refer to the term proportional to $n^i_{\rm s}v_i$ as `Doppler lensing'.
In order to model statistical observables (e.g. correlation functions) on small-scales, it is acceptable to 
replace the two-dimensional Laplacian $\nabla_\perp^2$ with the three-dimensional one $\nabla^2$, so that $\kappa$ can be expressed 
in terms of the matter overdensity using the Poisson equation. In fact, 
the effective lensing weight, $(\chi_{\rm s}- \chi)\,\chi/\chi_{\rm s}$, varies on  scales comparable to the Hubble radius and the line-of-sight integral that defines $\kappa$ heavily suppresses the contribution of radial Fourier modes with smaller wavelengths \citep[e.g.][]{Kaiser1992}. 
Although this approximation has been implemented to produce full-sky mock catalogues \citep[e.g.][]{Fosalbaetal2008,Fosalbaetal2015}, 
we do not use it since we want to study galaxy clustering at wide angular separations. 
 
\subsubsection{Link with N-body simulations} 
To evaluate $\delta x^\mu$, $\delta \chi$ and ${\cal M}$, we need to compute the gravitational potentials appearing in equation~(\ref{New}) as a function of space and time. 
Since we want to apply our results to simulations, we need to derive the potentials starting from the particle distribution in the computer models. 
This corresponds to using the matter density contrast in the synchronous comoving gauge, i.e. $\delta_{\rm sim}\equiv\delta_{\rm syn}$. 
Fortunately, to linear order in the perturbations and for a pressureless fluid in a universe with $\Lambda$CDM background,  
the source equation for $\Psi$ in the Poisson gauge can be re-written in terms of $\delta_{\rm syn}$ 
as the standard Poisson equation \citep[e.g.][]{Chisari:2011iq, Green-Wald2012}. 
Therefore, the complete dictionary we use to translate from the simulations to the Poisson gauge is: 
\begin{equation} \label{eq_poission} 
 \Phi=\Psi=\phi\;, \ \ \  \nabla^2 \phi=4 \pi G a^2 \bar\rho_{\rm m}  \delta_{\rm sim}\;, \   {\rm and} \  v^i=v^i_{\rm sim}\;, 
\end{equation} 
where $G$ denotes Newton's gravitational constant and $\bar\rho_{\rm m} $ is the matter density in the FRW background.

\subsection{Light cones from simulations} 
\label{sect_lightcones} 
In this section we explain the numerical methods we use to implement the theory discussed above and build mock light cones starting from the output of a simulation. 
We begin with the calculation of the gravitational potential. Following a standard procedure, 
we use the particle distribution in each snapshot to compute the matter density contrast on a regular Cartesian grid with the cloud-in-cell method \citep[e.g.][]{HockneyEastwood1988}. 
We then solve the Poisson equation using a fast Fourier transform and obtain $\phi({\bf x}_{\rm r},t)$ as well as its spatial derivatives (by spectral differentiation). 
Partial time derivatives of the potential are computed with a finite-difference method that combines several consecutive snapshots at fixed comoving position ${\bf x}_{\rm r}$. 
 
\begin{figure}
 \includegraphics[width=0.8\textwidth,bb=100 0 1200 550,keepaspectratio=true]{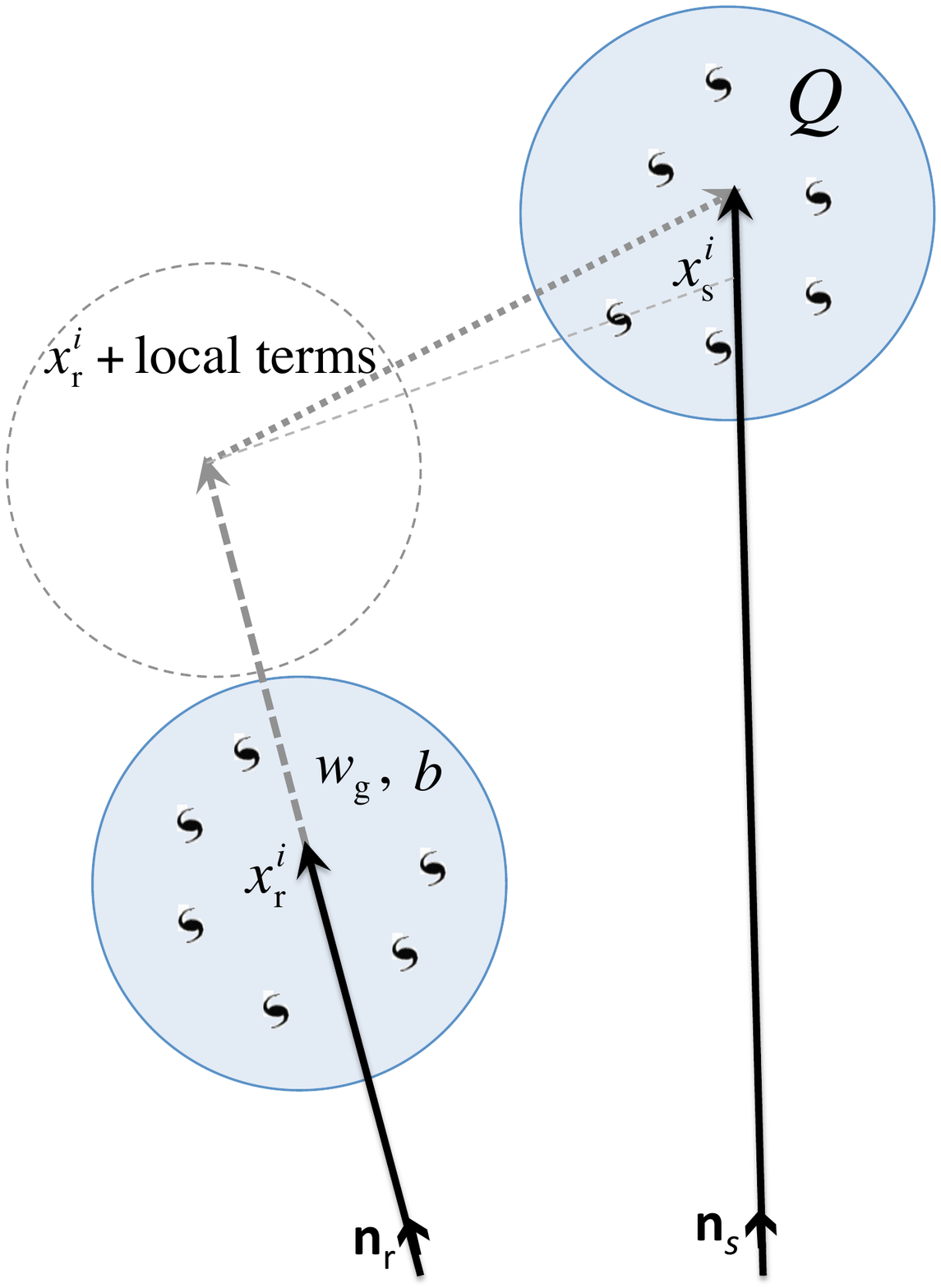}
 \caption{Schematic summarising how galaxies (Section~\ref{sect_lightcones}) or N-body particles (Section~\ref{sect_galaxy_dist}) are shifted to build the light cones.
We first apply the correction due to local terms (dashed arrow) and then compute the shift produced by the non-local contributions (dotted arrow).
In runs with low mass resolution, each particle `contains' $w_{\rm g}$ galaxies.
 The linear bias coefficient $b$ and $w_{\rm g}$ are calculated at the real-space position $x_{\rm r}^i$
 while the magnification bias $Q$ is computed at the redshift-space position $x_{\rm s}^i$.
 }
 \label{pic_shift}
\end{figure}

There are a few subtleties at play in the calculation of the galaxy shift and the magnification given in equations~(\ref{eq_delta_los}), (\ref{eq_delta_len}) and (\ref{eq_magni}).  
All these quantities include local terms evaluated at a specific position and non-local parts that are expressed as integrals along the line of sight to the observer.  
The integrals should be taken in redshift-space where the photon path is a straight line.  
Since $|\phi|\ll 1$ and deflections are generally small, we take the integrals in real space which is correct to linear order in the perturbations as in the Born approximation in quantum mechanics. 
We use the fast voxel traversal algorithm by \citet{Amanatides-Woo} to perform the integrals within the grid over which the gravitational 
potential is evaluated. All functions appearing in the integrands are interpolated in time (here converted into the line-of-sight distance to make sure that everything is computed on the backward light cone of the observer) such that their first derivatives are continuous. 
Note that the integration path starts at the observer and ends at a fixed redshift-space position which is unknown for all the simulated galaxies. 
Although to linear accuracy we could use the real-space position of the galaxies, we implement the following procedure which is slightly more accurate (see also Fig.~\ref{pic_shift}).  
(i)  We evaluate all local terms in equations~(\ref{eq_delta_los}) and (\ref{eq_delta_len}) and shift the galaxies accordingly. 
(ii) The non-local terms are estimated with integrals that run from the observer to the position of the galaxy shifted by the local terms. 
(iii) An additional shift due to the non-local terms is imposed to obtain the final redshift-space position. 
In principle, steps (ii) and (iii) could be iterated until numerical convergence is achieved.  
However, this is not necessary in practice since the local terms generate much larger shifts than the non-local ones.  
This is fortunate because the integration along the line of sight is by far the slowest element of the LIGER code. 
Magnification is computed along the same lines \citep[similar line-of-sight integrations have been used by][ for weak-lensing studies]{WhiteHu2000,Kiesslingetal2011,Fosalbaetal2008,Fosalbaetal2015}. 
 
Of course we do not shift all the galaxies at all times as this would significantly slow down the code and also be useless. We first identify the snapshots  
within which a given galaxy would cross the backward light cone of the observer in the absence of metric perturbations. 
We then calculate and apply the redshift-space displacements considering a few outputs surrounding this time. 
Finally, we compute the intersection of the world line of the galaxy with the straight light cone of the observer in redshift space and we save this position and the corresponding magnification.  
 
Each light cone identifies a sub-region in space-time corresponding to a three-dimensional ball in comoving redshift space. 
Its radius is limited by the box size of the underlying simulation. 
To avoid replications or spurious correlations due to the periodic boundary conditions applied in cosmological simulations, we limit the radius of the balls to one third of the box size. 
This way each light cone covers nearly 15 per cent of the simulation volume and we can place five different observers from a single run avoiding intersections.  

\subsection{Light cones based on dark-matter-only simulations}\label{sect_galaxy_dist} 
The LIGER method is general and can be used with all kinds of cosmological simulations.  
Whenever galaxy positions and luminosities are available (from either a hydrodynamic simulation or a semi-analytic model based on an N-body run), it is straightforward to apply shifts and magnifications at their locations.  
However, the transverse size of the light cones rapidly increases with redshift so that very large simulation boxes are required to cover 
wide opening angles. In this case, running simulations with sufficient spatial and mass resolution to follow galaxy formation is computationally challenging.  
For this reason, LIGER has the option to shift the dark-matter particles themselves and create the galaxy density field a posteriori. 
The problematic step is to account for galaxy biasing. 
We illustrate how our implementation works by reasoning in terms of 
continuous densities. 

To linear order in the perturbations, we can write the matter density contrast in redshift space as 
\begin{eqnarray}\label{eq_delta_rsd} 
 \delta_{\rm s}=\delta_{\rm sim}+\delta_{\rm RSD}\;, 
\end{eqnarray} 
where we have conveniently collected the corrections
due to the metric distortions into the term \citep{Yoo:2008tj, Yoo:2009au, Bonvin:2011bg, Challinor:2011bk, Jeong:2011as} 
\begin{eqnarray}
\label{deltaRSD}
\delta_{\rm RSD}\!\!\!\!\!&=&\!\!\!\!\!-\left(\frac{\partial_0\cH}{\cH^2}+\frac{2}{\chi_{\rm s} \cH}\right)\delta \ln a +\Psi_{\rm e} -2\Phi_{\rm e}+\frac{(\p_0\Phi)_{\rm e}}{\cH}+3 \cH \phi_v \nonumber\\
&\,&\!\!\!\!\!-\frac{1}{\cH} \left[n^i_{\rm s}\p_i \left(n^j_{\rm s} v_j\right)\right]_{\rm e}
+\frac{2}{\chi_s} \int_0^{\chi_s} \left(\Phi+\Psi \right)\ud \chi -2\kappa\;,
\end{eqnarray}
in which $\phi_v$ is the linear velocity potential\footnote{This term originates because $\delta_{\rm sim}$ is defined in the synchronous comoving gauge while all the rest is set in the Poisson gauge.}
at the galaxy position (i.e. $v_i=\partial_i\phi_v$)
and the apparent redshift change $\delta \ln a=\delta z/(1+z)$ due to the perturbations is 
\begin{eqnarray} 
\delta \ln a  = \Psi_o-(n_{\rm s}^i  v_{i})_o  - \Psi_{\rm e} + (n_{\rm s}^i  v_i)_{\rm e}- \int_0^{\chi_s} \p_0\left(\Phi+\Psi \right) \ud \chi \; .
\end{eqnarray} 
To the same accuracy, galaxy clustering in redshift space can be modelled  
in terms of three redshift-dependent bias parameters, $b$, ${\cal Q}$ and ${\cal E}$, encoding information about different properties of the galaxy population under study, namely \citep{Challinor:2011bk, Jeong:2011as} 
\begin{eqnarray} 
\delta_{\rm g, s}=b\,\delta_{\rm sim}+{\cal Q}\,({\mathcal M}-1)+{\cal E}\,(\delta \ln a-\cH\phi_v)+\delta_{\rm RSD}\;. 
\label{galaxydistortion} 
\end{eqnarray} 
The expression above assumes that the intrinsic perturbation in the galaxy number density is $\delta_{\rm g, r}=b\,\delta_{\rm sim}$ 
with $b$ the linear bias parameter. 
It also considers that lensing magnification alters the observed number density of galaxies. This effect is quantified by the magnification-bias parameter  
\begin{eqnarray} 
 {\cal Q}=-\frac{\partial\ln \bar{n}_{\rm g}}{\partial\ln L}\Big|_{L=L_{\rm lim}}, 
\end{eqnarray} 
where $\bar{n}_{\rm g}(>L)$  denotes the comoving number density of galaxies with luminosity larger than $L$ 
and the derivative is evaluated at the (redshift-dependent) limiting luminosity of the survey.\footnote{For 
simplicity, we assume that the list of targets for spectroscopic observations is flux limited. In case also a size cut 
is applied, another redshift-dependent function should be added to ${\cal Q}$ since gravitational lensing also alters the size of galaxy images \citep{Schmidtetal2009}.}  
Finally, equation (\ref{galaxydistortion}) takes into account that the comoving number density of galaxies in the sample might change with redshift. This phenomenon is described by the `evolutionary bias' parameter 
\begin{eqnarray} 
{\cal E}=-\frac{\partial \ln \bar{n}_{\rm g}}{\partial \ln (1+z)}\;.
\end{eqnarray} 

Our goal is to connect $\delta_{\rm g, s}$ with the particle density in the N-body simulations.
For simplicity, we assume that $|{\cal H}\phi_v|\ll|\delta\ln a|$ and neglect the velocity potential which could influence galaxy clustering only on scales comparable with the Hubble radius.
Then equations (\ref{eq_delta_rsd}) and (\ref{galaxydistortion}) give
\begin{eqnarray} 
 \delta_{\rm g, s}=(b-1)\delta_{\rm sim}+\delta_{\rm s}+{\cal E}\,\delta \ln a+{\cal Q}\,({\mathcal M}-1)\;,
\label{emphdeltag} 
\end{eqnarray} 
which can be used to derive the local galaxy number density $n_{\rm g,s}=\bar{n}_{\rm g}\,(1+\delta_{\rm g,s})$. 
By expressing the matter fields in terms of the density of N-body particles, i.e. 
$\delta_{\rm sim}+1={n_{\rm sim,r}}/{\bar{n}_{\rm sim}}$ and $\delta_{\rm s}+1={n_{\rm sim,s}}/{\bar{n}_{\rm sim}}$, we can write
\begin{eqnarray}\label{eq_numbercount1}
 n_{\rm g,s}\!\!\!\!\!&=&\!\!\!\!\!(b-1)\left(w_{\rm g}n_{\rm sim,r}-\bar{n}_{\rm g}\right)+w_{\rm g}n_{\rm sim,s}+\bar{n}_{\rm g}{\cal E}\,\delta \ln a\nonumber\\
 &&\!\!\!\!\!+\bar{n}_{\rm g}{\cal Q}({\cal M}-1),
\end{eqnarray}
where $w_{\rm g}=\bar{n}_{\rm g}/\bar{n}_{\rm sim}$ denotes the mean number of galaxies per simulation particle at a given redshift.
The products $w_{\rm g}n_{\rm sim,r}$ and $w_{\rm g}n_{\rm sim,s}$ rescale the unbiased density fluctuations in the simulations to the galaxy mean density.
The magnification term in equation~(\ref{eq_numbercount1}) reflects the relative change of the galaxy counts per particle which is proportional to $w_{\cal Q}={\cal M}^{\cal Q}$. 
For $|{\cal M}-1|\ll1$, we can thus write $w_{\rm g}w_{\cal Q}n_{\rm sim,s}=w_{\rm g}[{\cal Q}({\cal M}-1)+1]n_{\rm sim,s}$, so that
\begin{eqnarray}
 n_{\rm g,s}\!\!\!\!\!&=&\!\!\!\!\!(b-1)\left(w_{\rm g}n_{\rm sim,r}-\bar{n}_{\rm g}\right)+w_{\rm g}w_{\cal Q}n_{\rm sim,s}+\bar{n}_{\rm g}{\cal E}\,\delta \ln a\,.
\end{eqnarray}
By using the definition of ${\cal E}$ and linearising, $w_{\rm g}(z)w_{\cal Q}(\hat{\bf n}_{\rm s},z)n_{\rm sim,s}(\hat{\bf n}_{\rm s},z)+\bar{n}_{\rm g}(z){\cal E}\,\delta \ln a$ coincides with $w_{\rm g}(\bar{z})w_{\cal Q}(\hat{\bf n}_{\rm s},z)n_{\rm sim, s}(\hat{\bf n}_{\rm s},\bar{z})$ where $\bar{z}=z-\delta z$ is the redshift in absence of perturbations
(note that to first order it is equivalent to evaluate $w_{\cal Q}$ at $z$ or $\bar{z}$).
Eventually, making explicit the arguments of all functions, we obtain 
\begin{eqnarray} 
 n_{\rm g, s}(\hat{\bf n}_{\rm s},z)\!\!\!\!\!&=&\!\!\!\!\![b(\bar{z})-1]\left[w_{\rm g}(\bar{z})\,n_{\rm sim, r}(\hat{\bf n}_{\rm r},\bar{z})-\bar{n}_{\rm g}(z)\right]\nonumber\\
 &\,&\!\!\!\!\!+w_{\rm g}(\bar{z})w_{\cal Q}(\hat{\bf n}_{\rm s},z)\,n_{\rm sim, s}(\hat{\bf n}_{\rm s},\bar{z})\,. 
\label{eq_numberdens} 
\end{eqnarray} 
We use this expression to compute $n_{\rm g, s}$ from the simulations (see Fig.~\ref{pic_shift} for a schematic representation).
In practice, we weigh the shifted and unshifted dark-matter particles according to equation~(\ref{eq_numberdens}).
Once the light cone for the matter has been constructed, it is very fast to build the galaxy mocks for many different galaxy populations. 
This corresponds to changing the functions $w_{\rm g}(z)$, $b(z)$ and ${\cal Q}(z)$. 

\subsubsection{Doppler terms}
Isolating the terms proportional to the velocity field in the right-hand side of equation (\ref{galaxydistortion}) we obtain
\begin{eqnarray}
 \delta_{\rm g, v}=-\frac{1}{\cal H} \left(\frac{\partial n_{\rm s}^iv_i}{\partial \chi_{\rm s}}\right)_{\rm e}-
\frac{ \alpha(\chi_{\rm s})}{\cH \chi_{\rm s}}
 \left[{\left(n_{\rm s}^iv_i\right)_{\rm e}-\left(n_{\rm s}^iv_i\right)_{\rm o}}\right] \;,
 \label{hamilton}
\end{eqnarray}
where $\partial/\partial\chi_{\rm s}=n^i\partial_i$,
\begin{equation}
\alpha(\chi_{\rm s})=\gamma_0+\gamma_1\, \cal H\chi_{\rm s}
\end{equation}
and, assuming a flat $\Lambda$CDM universe,
\begin{equation}
\label{alphaterms}
\gamma_0=2(1-{\cal Q})\ \ \ \ \ {\rm and} \ \ \ \ \ 
\gamma_1=1-\frac{3}{2} \Omega_{\rm m}(z)-{\cal E}+ 2 {\cal Q}\;.
\end{equation}
Equation (\ref{hamilton}) coincides with the seminal result for the linear redshift-space distortions derived by \citet{Kaiser:1987qv}. 
In the classical literature, the function $\alpha$ is often written as $\alpha=2+{\partial  \ln \bar{n}_{\rm g}}/{\partial \ln \chi_{\rm s}}$
\citep{Kaiser:1987qv, Zaroubi-Hoffman1996,1998HamiltonReview}. 
Taking into account that we observe galaxies on our past light-cone reveals that several physical effects influence $\alpha$
\citep[see also][]{McDonald2009, Yoo2009, Bertacca:2012tp, Raccanelli:2016avd}.
Equations (\ref{alphaterms}), get contributions from
 geometric distortions, redshift evolution (or redshift-dependent selection effects),  
Doppler magnification and cosmic acceleration.

Following a standard practice in cosmology, we label the expression proportional to $\alpha$  
in equation (\ref{hamilton}) with the collective name of Doppler terms. 
Their contribution is usually neglected in clustering studies.
In fact, for an ideal galaxy sample with $\alpha\simeq 2$ and if the depth of a galaxy redshift survey is much larger than the comoving wavelength of interest, the Doppler induced $\delta_{\rm g,v}$ is heavily suppressed 
(due to the $\chi_{\rm s}^{-1}$ scaling) with respect to the signal generated by the radial velocity gradient which is always comparable to density perturbations \citep{Kaiser:1987qv}. 
This reasoning relies upon the distant-observer approximation. 
However, it has been shown that the Doppler corrections can alter the galaxy autocorrelation function at large angular separations in a significant way
\citep{Papai:2008bd, Raccanelli:2010hk}. In this case, the two terms on the right-hand side of 
equation (\ref{hamilton}) can be of comparable sizes. Moreover, $Q$ and ${\cal E}$ can drive $\alpha$ sensibly away from 2 and thus enhance
the chance of detecting the Doppler terms from observational data. We will return to this issue in Section \ref{sect_lowred}.
 
\section{Examples} 
\begin{figure*}
 \includegraphics[width=\textwidth,bb=0 0 1032 1048,keepaspectratio=true]{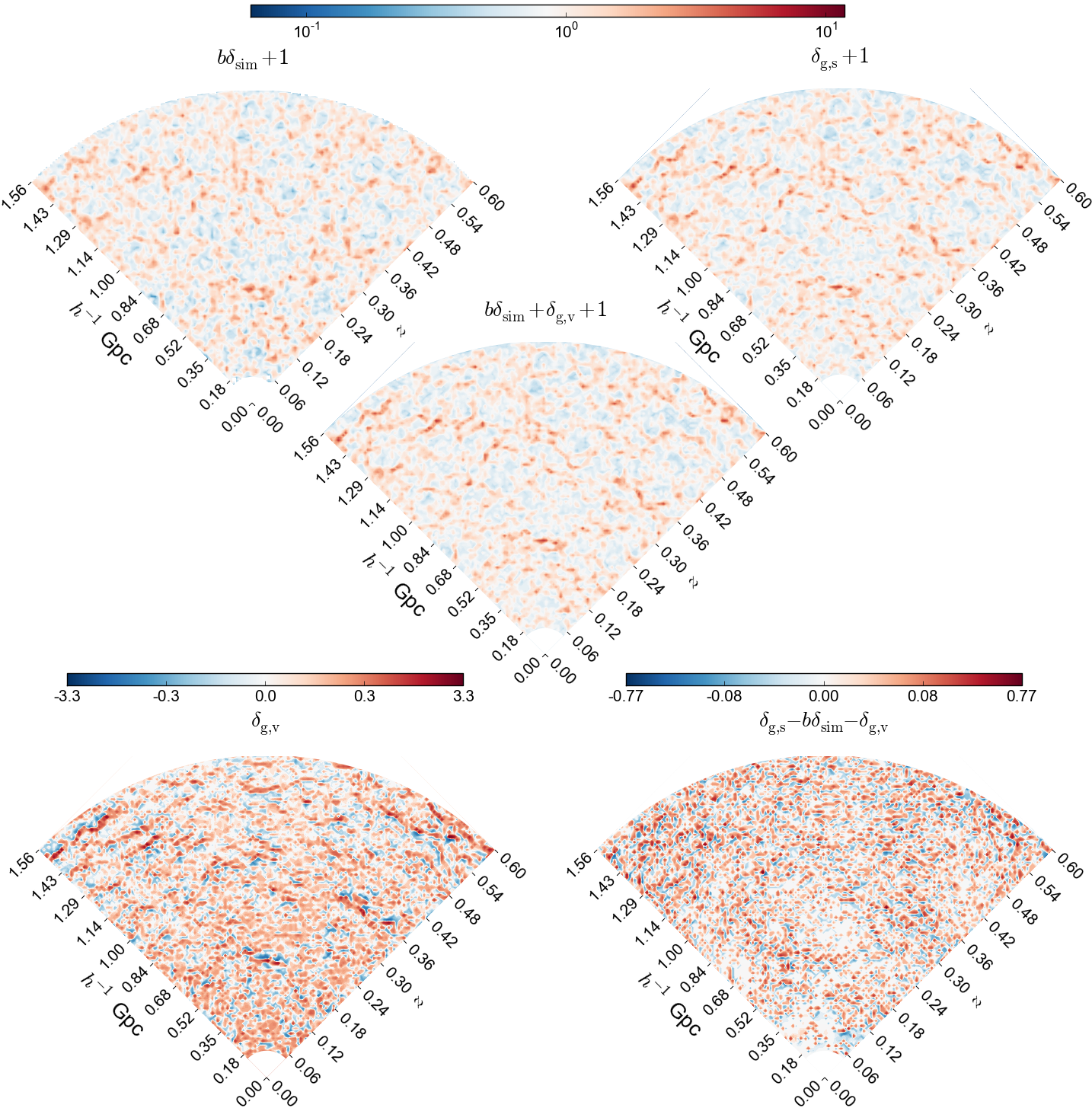}
 \caption{Wedge plots extracted from a sample like cone. 
The top images show $b\delta_{\rm sim}$, $\delta_{\rm g,s}$, and $b\delta_{\rm sim}+\delta_{\rm g,v}$.
In all cases,  the maps show the galaxy density contrast projected onto the plane of the page within a slice of 
5 arcmin thickness. The observer is located at the vertex of the wedge and the labels indicate redshift and comoving distance.
The bottom panels highlight the differences between the various density fields (note the symmetric log-scale).}
 \label{pic_stripe}
\end{figure*}

We present two sample applications of LIGER:
first, we estimate the importance of magnification bias in an Euclid-like survey and 
then we investigate the detectability of Doppler terms in a low-redshift galaxy catalogue based on the concept of the Square Kilometre Array (SKA).
To begin with, we introduce the numerical simulations and the statistical methods we use. 
We then describe the specifications for the surveys and discuss our results in 
Sections~\ref{sect_euclidlike} and \ref{sect_lowred}.

\subsection{N-body simulations}
\label{Nbodysims}
We run a large number of cosmological N-body simulations using the L-PICOLA code \citep{2015HowlettLPicola} and subsequently apply LIGER to their outputs.
L-PICOLA is an implementation of the COLA method \citep{2013TassevCola} in which the large-scale dynamics is solved using second-order Lagrangian perturbation theory while a particle-mesh algorithm is used for the
small scales.
This technique is orders of magnitude faster than standard N-body codes and accurately simulates
the clustering of matter on large scales. 
This makes it an ideal tool to build large mock catalogues for studying galaxy clustering although
it does not resolve the internal dynamics of dark-matter haloes.

Our simulations include $1024^3$ particles in a periodic cube with side length $L$.
In order to cover the relevant volumes, we use very large values of $L$, namely $12 h^{-1}$ Gpc for our first application and $5 h^{-1}$ Gpc for the second one.
This way we obtain 165 light cones extending to redshift 2.3 for the Euclid-like mocks and
125 light cones extending to redshift 0.6 for the more local mocks (an example is shown in Fig.~\ref{pic_stripe}).
In all cases, the gravitational potential in LIGER is evaluated on a grid with $512^3$ cells.

It is worth stressing that LIGER is completely general and can be applied to the output of any N-body code.
Here we use L-PICOLA because it is ideal for our purposes. 
Note that we do not make use of the built-in feature to build light cones on the fly implemented in L-PICOLA.
However LIGER could be merged with it in the future.

\subsection{Angular power spectra}
Our examples focus on large-scale galaxy clustering that we quantify in terms of the angular power spectrum.
We first divide our mock light cones into multiple redshift bins and measure the projected galaxy number density contrast 
on the sky, $\sigma_{\rm g}^{(i)}(\btheta)$, for each of them (labelled by the index $i$).
We then decompose $\sigma_{\rm g}^{(i)}(\btheta)$ in spherical harmonics, $\sigma_{\rm g}^{(i)}(\btheta)=a_{lm}^{(i)}\, Y_{lm}(\btheta)$, with
\begin{equation}
 a_{lm}^{(i)}=\int{d}^2\btheta\, \sigma_{g}^{(i)}(\btheta)\,Y_{lm}^\ast(\btheta)\;,
\end{equation}
and measure the angular auto- and cross-spectra between all redshift bins using
\begin{equation}
 C_{l}^{(ij)}=\frac{1}{2l+1}{\sum_{m=-l}^{l} a_{lm}^{(i)}\,a_{lm}^{(j)\ast}}\;.
 \end{equation}

In practice, we use the Healpix algorithm \citep{2005GorskiHealpix}  to build digitized maps of $\sigma_{\rm g}^{(i)}(\btheta)$ from which we calculate the power spectra.
In all cases, we make sure that the pixel size does not affect our results in the range of scales of interest.
The galaxy density in a pixel is computed from the distribution of the N-body particles as described in Section~\ref{sect_galaxy_dist}.
Since our particles are rather massive ($1.2\times10^{14}\,h^{-1}M_\odot$ for the $12\,h^{-1}$ Gpc boxes and
$8.8\times10^{13}\,h^{-1}M_\odot$ for the $5\,h^{-1}$ Gpc ones),
each of them `contains' multiple galaxies (i.e. $w_{\rm g}\sim 10$).
This is unavoidable given the extremely large volumes covered by our simulations and the obvious limitations in computing time and memory usage.
Although the resulting overdensity field has increased shot noise with respect to the actual galaxy distribution, 
our statistical analysis is not influenced by it. In fact, we never attempt to subtract shot noise from the power spectra as our study is based on the comparison
of different sets of mock catalogues. 
Even more importantly, in all cases, shot noise is by far subdominant with respect to the sample and cosmic variance of the clustering signal we are interested in (note that
our main results are based on the analysis of cross statistics between galaxy samples at different redshift). 

We also take into account that most survey geometries do not cover the full sky and exclude wide regions surrounding the galactic plane.
In order to simulate a realistic setting, we mask an appropriate amount of the sky around the equator of the observer and consider two distinct regions
around the poles covering a fraction $f_{\rm sky}$ of the celestial sphere.
Spherical harmonics are not orthogonal over finite solid angles and
the pseudo power spectrum $\tilde{C}_{l}^{(ij)}$ measured from the cut sky does not coincide with $\hat{C}_{l}^{(ij)}$ \citep{Peebles1973,Wandeltetal2001}.
We use the standard method by \citet{2002Hivon} to construct an unbiased estimate of the full-sky spectra $\hat{C}_{l}^{(ij)}$ which is obtained multiplying
$\tilde{C}_{l}^{(ij)}$ by the inverse of a mode-mode coupling matrix that depends on the survey mask.

\subsection{Statistical analysis}\label{sect_significance}
Our sample applications aim at quantifying the detectability of some specific redshift-space effects from measurements of two-point statistics.
We assume that we can perfectly model the different contributions to the clustering signal and check whether including or excluding some
of them improves or worsen the fit to the mock data including the full physics.
In particular, we proceed as follows.
We isolate a particular effect (say, e.g., magnification bias) and denote its expected partial contribution to the model galaxy power spectrum as $C_l^{\rm (A)}$ so that
its complementary part is $C_l^{\rm (B)}=C_l-C_l^{\rm (A)}$ (here the superscripts $(ij)$ indicating the redshift bins are understood to simplify notation).
We then fit the power spectra extracted from our mock catalogues with the model
$M_l=\epsilon C_l^{\rm (A)}+C_l^{\rm (B)}=C_l+(\epsilon-1)C_l^{\rm (A)}$ where the coefficient $\epsilon$ 
can only assume the values zero or one. The question we want to address is to what statistical significance the data favour $\epsilon=1$, i.e.
how necessary it is to add $C_l^{\rm (A)}$ to the model in order to fit the data $\hat{C}_l$ .
All this boils down to comparing the quality of the fit obtained using $\epsilon=0$ and $\epsilon=1$.
This exercise can be performed following different statistical procedures which give very similar results. 

\subsubsection{Frequentist approach: simple hypotheses (SH)}\label{sect_significance_simple}
We want to test the null hypothesis $H_0:\epsilon=0$ against the alternative hypothesis $H_1:\epsilon=1$.
Let ${\cal L}_0$  and ${\cal L}_1$ denote the likelihood of the data under $H_0$ and $H_1$, respectively. 
Based on the Neyman-Pearson lemma, the likelihood-ratio statistic $\lambda={\cal L}_0/{\cal L}_1$
provides the most powerful test for two simple hypotheses.
If the data do not support $H_0$, then the likelihood ratio should be small. 
Thus, we reject the null hypothesis with confidence level $\tau$,
 if $\lambda\leq \omega$, where $\omega$ is a constant such that the probability $P(\lambda\leq \omega| H_0)=\tau$.
In order to determine $\tau$ it is thus necessary to determine the probability distribution of the test statistic under $H_0$.
Working with the log-likelihood $\chi^2=-2 \ln {\cal L}$, the rejection condition becomes $\Delta \chi^2=\chi_1^2-\chi_0^2>\ln \omega$.
Assuming Gaussian errors for the angular power spectrum with covariance matrix $\Sigma_{lm}=\langle \hat{C}_l \hat{C}_m\rangle-
\langle \hat{C}_l\rangle \langle \hat{C}_m\rangle$, 
we find that, under $H_0$,  $\Delta \chi^2$ follows a Gaussian distribution with
mean 
\begin{equation}
\mu=C_l^{\rm (A)} \Sigma^{-1}_{lm} C_m^{\rm (A)}
\end{equation}
and variance $4\mu$ (see Appendix \ref{Appendix}).
Therefore, we reject $H_0$ at the 95 per cent confidence level if $\Delta \chi^2>\mu+3.29\sqrt{\mu}$.
The coefficient 3.29 should be replaced with 4.652 to get a 99 per cent confidence level.
A formal $5\sigma$ rejection is obtained for $\Delta \chi^2>\mu+10\sqrt{\mu}$.

It is worth noticing that,
if the covariance matrix of the measurements does not depend on $\epsilon$, then the expected value of $\lambda$ under $H_1$ is
$E(\lambda | H_1)=-E(\lambda | H_0)=-\mu$. 
Therefore,  the mean values $E(\lambda | H_1)$ and $E(\lambda | H_0)$ will be separated by more than ${\cal N}$ standard deviations of the $\lambda$ distribution
only if $\sqrt{\mu}>{\cal N}$. This is why $\sqrt{\mu}$ is often denoted as the signal-to-noise ratio, $S/N$, of $C_l^{\rm (A)}$.
On the other hand, if the covariance depends on $\epsilon$, then  $E(\lambda | H_1)\neq-E(\lambda | H_0)$ and the dispersions around the mean of $\lambda$ under $H_0$ and $H_1$ will be different (see Appendix \ref{Appendix}).

\subsubsection{Frequentist approach: composite hypothesis (CH)}
We also consider a generalized likelihood-ratio test with a compound alternative hypothesis.
In this case we contrast the null hypothesis $H_0:\epsilon=0$ with $H_1:\epsilon \neq 0$.
Let ${\cal L}_{\rm max}$ be the maximum value of the likelihood of the data when $\epsilon$ is varied (between 0 and 1)
and ${\cal L}_0$ the corresponding likelihood under the null hypothesis.
We form the ratio $\lambda={\cal L}_0/{\cal L}_{\rm max}$ which is always between 0 and 1.
We assess the statistical significance of the test by comparing $\Delta \chi^2=-2 \ln \lambda$ to the chi-square distribution with one degree of freedom
(as we only tuned one parameter to determine ${\cal L}_{\rm max}$). Basically we convert $\Delta \chi^2$ into the corresponding percentile of the chi-square distribution. This is the confidence level at which the null hypothesis can be rejected.
In this case, 95 (99) per cent confidence corresponds to a critical value of $\Delta \chi^2=3.84$ (6.64).

\subsubsection{Fisher information}
An alternative approach consists of
quantifying the Fisher information that the angular power spectrum carries about $\epsilon$ (which, in this case, is assumed to be a real number).
Assuming Gaussian errors, we obtain that the Fisher `matrix' for $\epsilon$ is 
\begin{equation}
F=\partial_\epsilon M_l\, \Sigma^{-1}_{lm} \,\partial_\epsilon M_m=C_l^{\rm (A)} \Sigma^{-1}_{lm} C_m^{\rm (A)}
\equiv  \mu
\end{equation} 
\citep[we adopt the `field' perspective as in][]{Carron2013}.
It follows from the Cram\`er-Rao inequality that $\mu^{-1/2}$ gives a lower bound for the expected uncertainty on $\epsilon$
 (i.e. the mean curvature of the likelihood function ${\cal L}(\epsilon)$ at its peak). 
 Therefore, $\sqrt{\mu}$ corresponds to the signal-to-noise ratio with which $\epsilon$ can be measured.

\begin{figure}
  \centering
  \includegraphics[width=0.48\textwidth,bb=0 0 694 719,keepaspectratio=true]{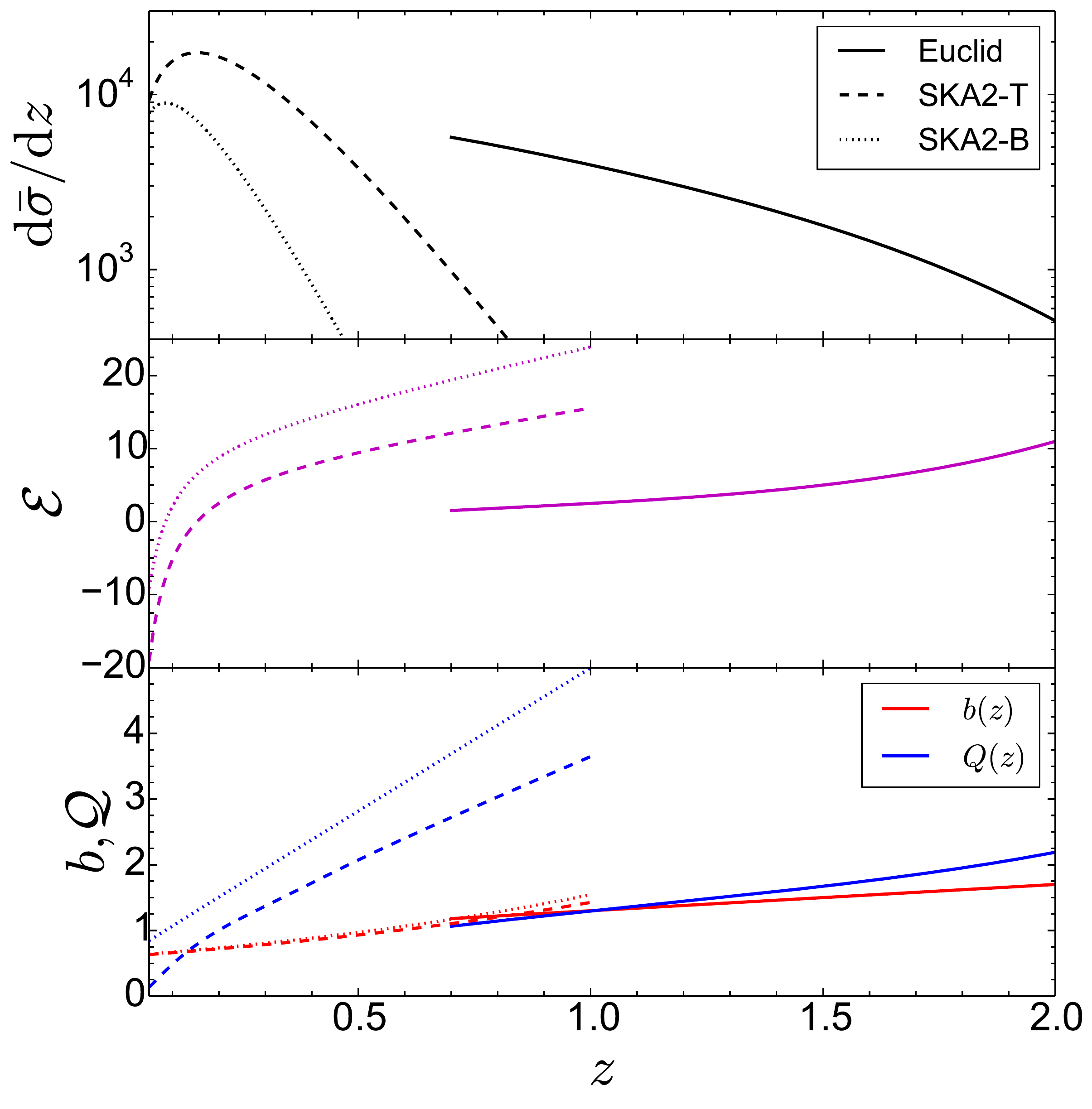}
  \caption{Expected characteristics of the Euclid (solid) and SKA2 redshift surveys. In the latter case we consider two flux sensitivities: 23 (dashed) and $60\,\mu{\rm Jy}$ (dotted).
  The mean galaxy number counts per square-degree and redshift (top),
  the evolutionary bias (middle), and the linear and magnification bias parameters (bottom) are shown as a function of redshift.}
  \label{Fig:surveydata}
\end{figure}

\begin{figure*}
  \centering
  \includegraphics[width=0.47\textwidth,bb=0 0 630 537,keepaspectratio=true]{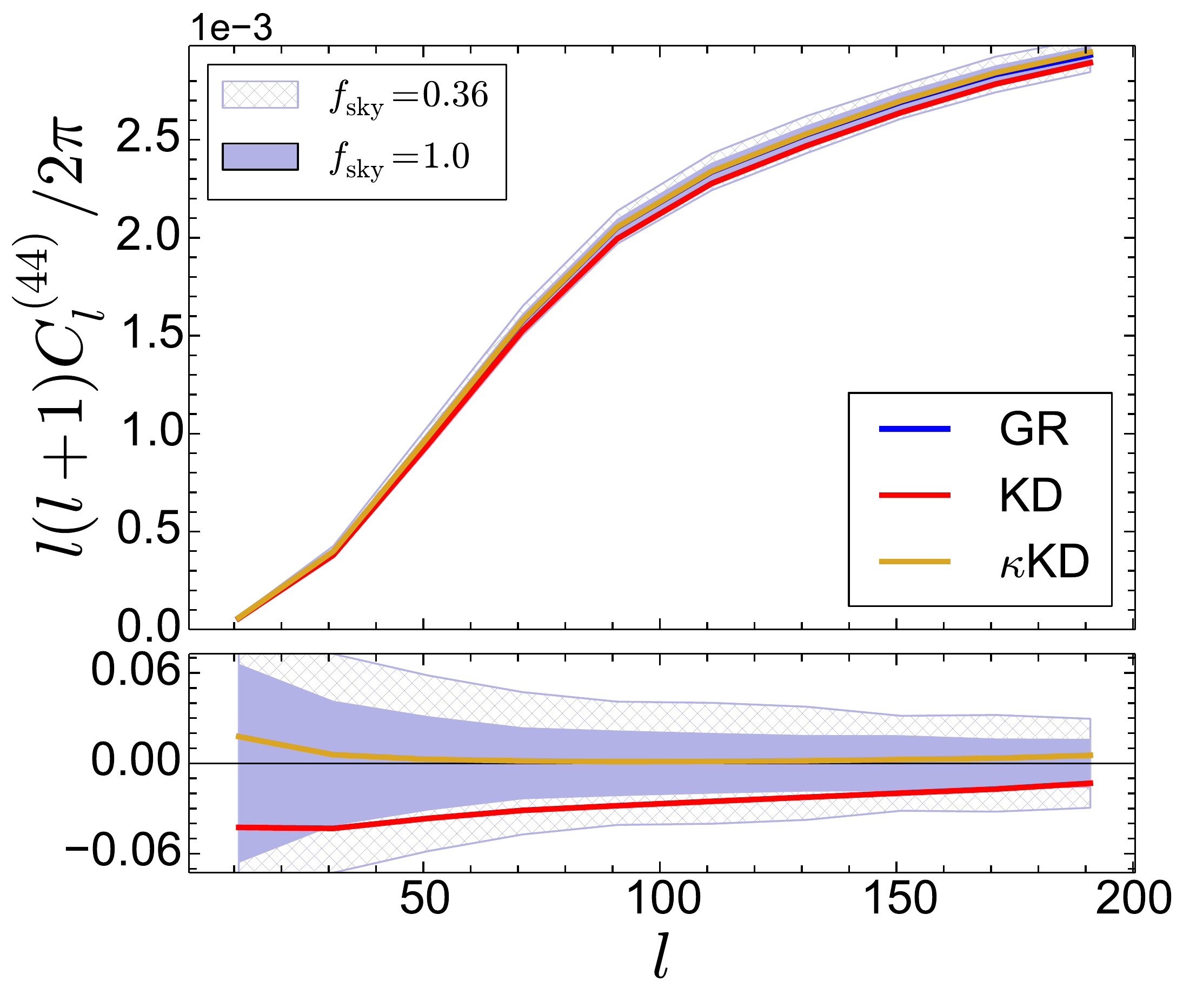}
  \hfill
  \includegraphics[width=0.47\textwidth,bb=0 0 634 537,keepaspectratio=true]{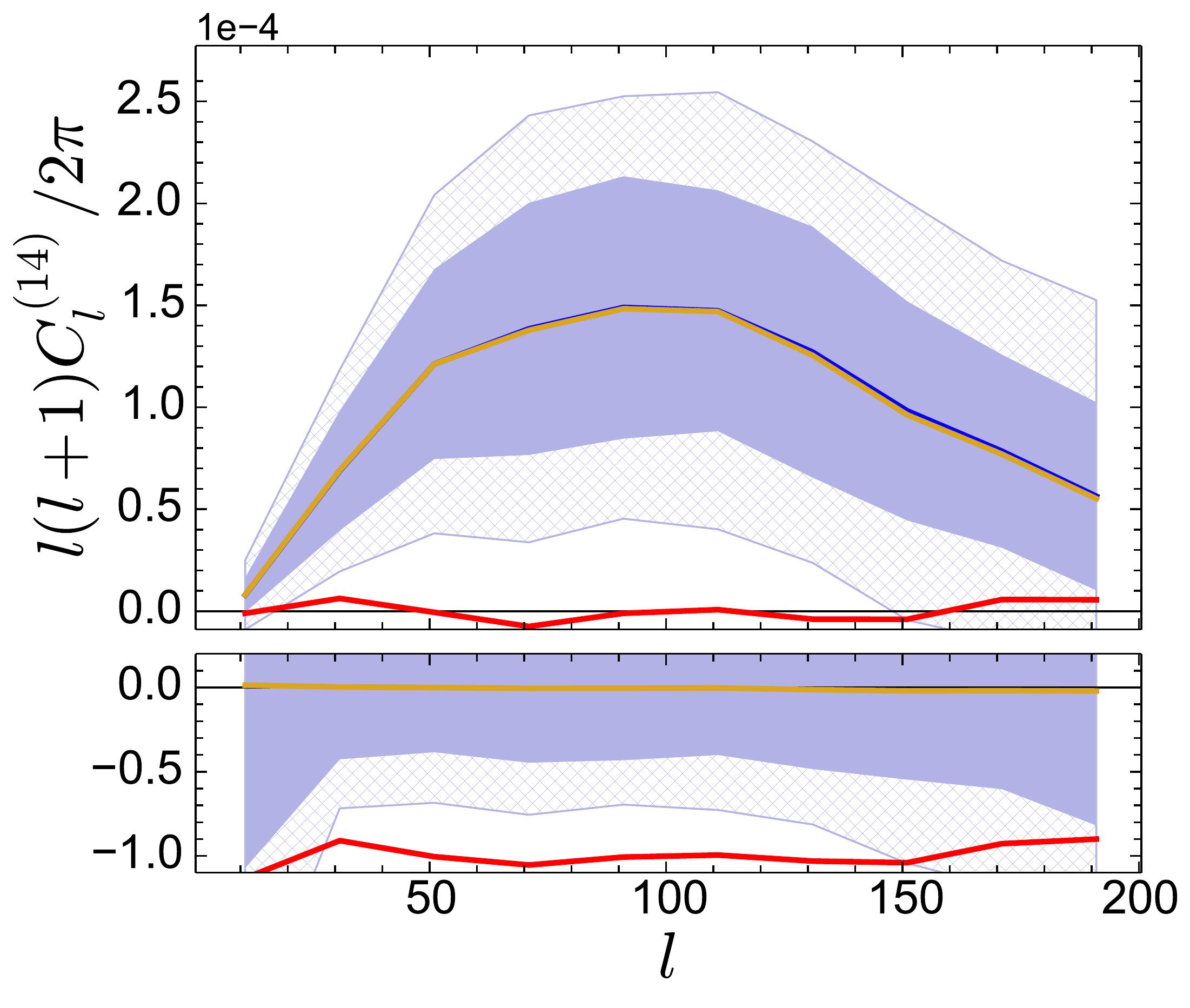}
  \caption{Angular power spectra for the galaxy distribution in a Euclid-like survey split into four equally populated redshift bins.
  The left-hand panel refers to the auto-spectrum of the highest-redshift bin while the right-hand panel shows the cross-spectrum between the lowest-and the highest-redshift bins. 
  The blue curves show the signal averaged over 165 mock catalogues that include relativistic effects to linear order in the perturbations (GR). 
  The shaded regions surrounding them indicate the standard deviation of the measurements for full-sky (light blue) and Euclid-like (cross pattern) mocks. 
  The red curves are obtained considering only the redshift-space distortions generated by the Kaiser and Doppler effects (KD) but ${\cal M}=1$.
  The yellow ones (which basically coincide with the blue ones) also consider magnification bias due to weak gravitational lensing.
  All spectra are averaged in bandpowers with $\Delta l=20$.
  The bottom panels highlight the relative difference of the KD and kKD models with respect to the GR signal.}
  \label{pic_angular_power}
\end{figure*}

\subsubsection{Bayesian approach (BA)}
Assuming a flat prior for $\epsilon$, the posterior probabilities for $\epsilon=0$ and $\epsilon=1$ are
$P_0=\frac{{\cal L}_0}{{\cal L}_0+{\cal L}_1}$ and 
$P_1=\frac{{\cal L}_1}{{\cal L}_0+{\cal L}_1}$,
respectively.
The Bayes factor, $K=P_0/P_1$, thus corresponds to the likelihood ratio ${\cal L}_0/{\cal L}_1$.
According to the Jeffrey scale, there is strong (decisive) evidence against $H_0$ if $K>10$ (100) which gives $\Delta \chi^2>4.605$ (9.21).
Using the alternative scale by Kass and Raftery gives very strong evidence for $\Delta \chi^2>10$.

\subsubsection{Likelihood estimation}
All the statistical methods listed above require the calculation of the likelihood function for $\epsilon$ in each mock realisation.
To do this, we need a model and the covariance matrix for the power spectra (we assume Gaussian measurement errors). 
We build an `exact' and unbiased model by averaging the angular power spectra obtained for all our mock light cones.
Since there are no particularly deviant realizations,  the average spectra are smooth.
In parallel, we use the maximum-likelihood estimator to get a first approximation for the covariance matrix, $\hat{\Sigma}_{lm}$.
It is well known that the precision matrix obtained by inverting $\hat{\Sigma}_{lm}$ is not very accurate.
Although we make sure to consider enough mock skies so that the covariance matrix of our data vector is invertible, it still contains considerable noise.
We thus use the shrinkage method \citep{CovShrinking2005} to reduce the noise in $\hat{\Sigma}_{lm}$.
As a target we use a diagonal matrix which is always compatible with our estimates.

\subsection{Magnification bias in a Euclid-like survey}\label{sect_euclidlike}
As a simple application of the LIGER method, we discuss the detectability of magnification bias in a Euclid-like survey.
Related work has been presented by \citet{DiDioetal2014} and \citet{Montanari:2015rga}. These authors focused on the Fisher information matrix as a forecasting tool,
while we base our study on the statistical analysis of a large number of mock catalogues.

\subsubsection{Euclid spectroscopic sample}
Euclid is a medium-class mission of the European Space Agency planned for launch in 2020.
It will map the distribution of star-forming galaxies through their redshifted H$\alpha$ emission
in the regions with galactic latitude larger than 30 degrees ($f_{\rm sky}=0.36$).
Low-resolution spectroscopy in the near infrared will be used to measure galaxy redshifts in the range $0.7<z<2.0$.

The specifics of the Euclid redshift survey depend on the poorly known properties of emission-line galaxies at moderate redshifts.
In order to calculate the redshift distribution of the galaxies as well as $Q(z)$ and ${\cal E}(z)$ we use
the redshift-dependent luminosity function by \citet[][model two]{Pozzettietal2016} and assume a limiting line flux of $3.0\times 10^{-16}$ erg cm$^{-2}$ s$^{-1}$.
Further, we use the linear fit for the galaxy bias $b(z)$ given in \cite{Pozzettietal2016}.
Our results are summarised in Fig.~\ref{Fig:surveydata}.

\begin{figure*}
 \includegraphics[width=0.945\textwidth,bb=0 0 1310 501,keepaspectratio=true]{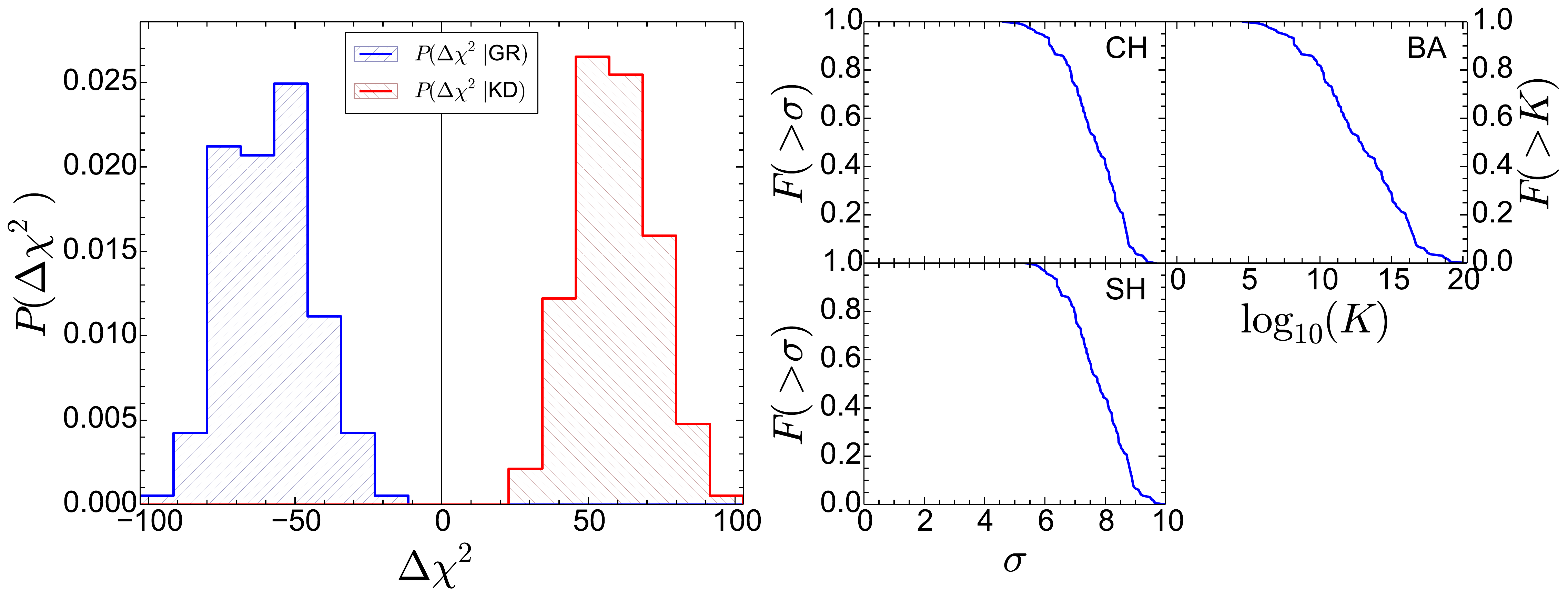}
 \caption{In order to explain the origin of the clustering signal, we consider two sets of Euclid-like mock catalogues (each one containing the same 165 skies) in which we arbitrarily switch on and off some effects.
 In the first group, redshift-space distortions are only generated by galaxy radial peculiar velocities (KD) whereas the second suite includes general-relativistic effects to linear order in the perturbations (GR). 
 We fit the angular power spectra $C_l^{(ij)}$ extracted from each mock catalogue with two models that have been obtained averaging the clustering signal within each series of catalogues.
 We then compute the change in $\chi^2$ for every sample.  
 In the left-hand panel, we compare the histogram of $P(\Delta\chi^2|{\rm GR})$ (on the left) versus $P(\Delta\chi^2|{\rm KD})$ (on the right). 
 The fact that the histograms are widely separated and do not overlap implies that an Euclid-like survey will clearly detect redshift-space distortions that are not included in the KD model..
 This is quantified in the right-hand panel where we plot the fraction $F(>x)$ of the GR mocks within which the KD model is rejected at a confidence level higher than $x$ using different statistical tests (for further details see Section~\ref{sect_significance}).}
 \label{pic_deltachi2_euclid}
\end{figure*} 

\begin{figure*}
 \includegraphics[width=0.945\textwidth,bb=0 0 1296 501,keepaspectratio=true]{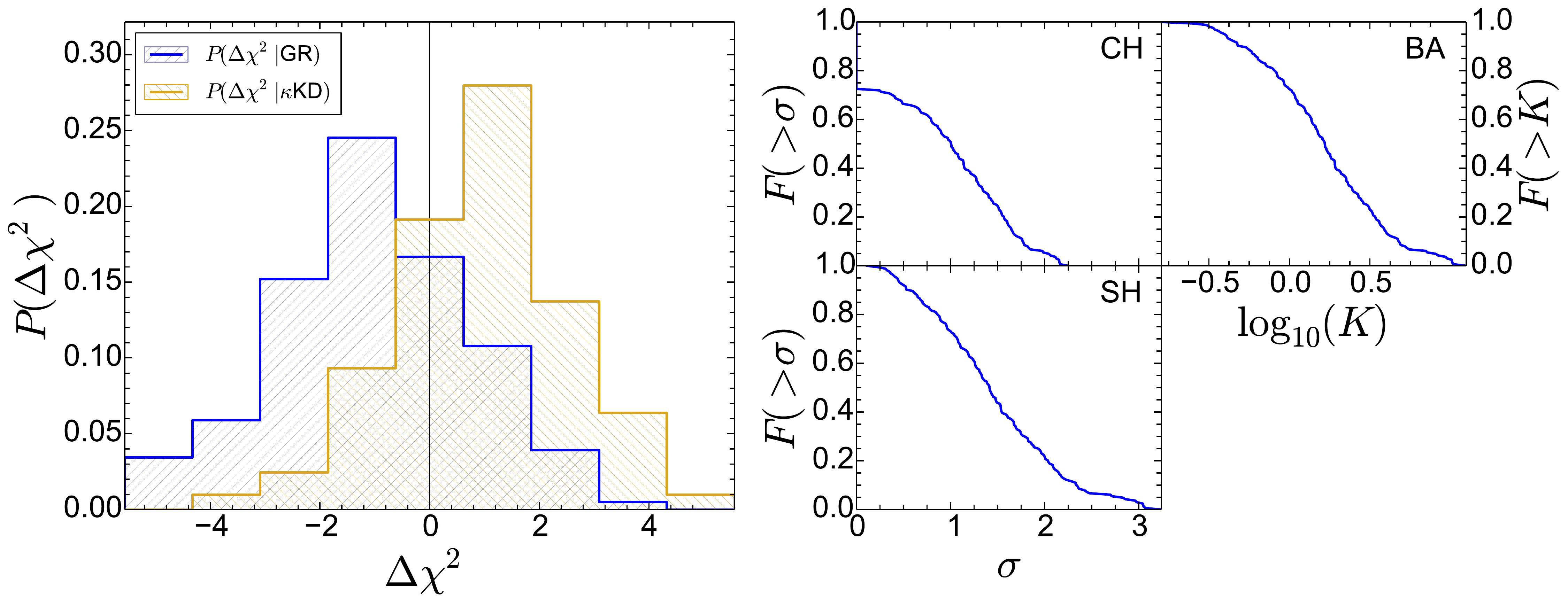}
 \caption{The same as Fig.~\ref{pic_deltachi2_euclid}, but for the $\kappa$KD model, i.e. including the effect of the convergence.
 For the CML test, 28 per cent of the GR mocks have $\Delta\chi^2>0$ and therefore favour the $\kappa$KD model.}
 \label{pic_deltachi2_euclid_kkd}
\end{figure*} 

\subsubsection{Angular power spectra}
We split each of the 165 Euclid mocks into four redshift bins (with boundaries $0.7, 0.86, 1.06, 1.35, 2.0$) that on average contain the same number of galaxies ($\sim1.6\times 10^7$).
We then compute the auto- and cross-power spectra among all bins. To reduce noise we do not resolve individual multipoles and use ten bandpowers with $\Delta l=20$.
Fig.~\ref{pic_angular_power} shows $C_l^{(44)}$ (left-hand panel) and $C_l^{(14)}$ (right-hand panel), where the indices 1 and 4 indicate the lowest
and highest redshift bin, respectively. 
The blue curves correspond to the mean power spectra extracted from the mock catalogues that include all redshift-space effects (hereafter GR).
Note that the auto-spectrum is more than ten times larger than the cross-spectrum. 
The shaded regions indicate 
the standard deviation of the spectra over the 165 realisations in the Euclid-like (cross pattern) and in the full-sky mocks (light blue).

It is interesting to analyse the different contributions to the spectra. The most commonly considered source of redshift-space distortions is 
the so-called Kaiser effect due to the gradient of the galaxy radial peculiar velocities, i.e. the first term on the right-hand side in equation (\ref{hamilton}).
In our approach, this correction derives from the $(n_{\rm s}^iv_i)_{\rm e}$ terms in the particle shift and is always mixed with the Doppler contribution.
In order to evaluate the relative importance of the velocity-induced shift,
we build a new set of Euclid mock catalogues (based on the same N-body simulations as the GR ones)
in which we replace equations (\ref{eq_delta_los}), (\ref{eq_delta_len}) and (\ref{eq:magnif}) with
$\delta \chi=-(n_{\rm s}^iv_i)_{\rm e}/\cH$, $\delta x^i=0$ and ${\cal M}=1$, respectively
(this is the standard way to implement redshift-space distortions in simulations and omits the terms proportional to ${\cal Q}$ in $\alpha$).
We dub these light cones KD, a short for `Kaiser and Doppler'.
The red curves in Fig.~\ref{pic_angular_power} show the mean clustering signal extracted from the KD mocks (which from now on we refer to as the KD model for the auto- and cross-spectra).
The lack of power with respect to the GR results is evident: $C_l^{(44)}$ is underestimated by $\sim 3$ per cent and  $C_l^{(14)}$ oscillates around zero.
In the left-hand panel of Fig. ~\ref{pic_deltachi2_euclid}, we demonstrate that these differences are highly significant. The histogram on the left-hand side displays the distribution
of $\Delta \chi^2=\chi^2_{\rm GR}-\chi^2_{\rm KD}$ obtained fitting all the $C_l^{(ij)}$ from the 165 GR mocks with the GR and the KD models, respectively. Similarly, the histogram on the right-hand side
shows the corresponding distribution for the KD mocks.  
Based on the fact that the histograms are well separated, we conclude that an Euclid-like survey should be able to
detect the signature of redshift-space distortions that are not included in the KD model.
To better quantify how inaccurately the KD model fits the mock GR data, we 
apply the statistical tests we have introduced in Section~\ref{sect_significance}. 
The cumulative distribution over the 165 GR mocks of the significance with which the KD model is rejected is shown in the right-hand panel of Fig.~\ref{pic_deltachi2_euclid}.
Typically the data disfavour the simpler model with 8$\sigma$ confidence or a Bayes factor of $10^{13}$.
The precise statistical significance of this result is very sensitive to the assumed ${\cal Q}(z)$.
For instance, it increases to 14$\sigma$ if we use model 3 from \cite{Pozzettietal2016}.
 
This result is not surprising. It is well known that weak gravitational lensing alters the observed clustering signal in deep magnitude-limited surveys \citep{Turner1980}.
The influence of lensing is twofold: 
(i) the actual magnitude limit of the survey fluctuates on the sky and with redshift;
(ii) the surface density of galaxies on the sky (and thus their volumetric density in redshift space) is changed.
The last two terms in equation (\ref{deltaRSD})
and the $Q$-dependent term in equation
(\ref{galaxydistortion}) summarise the net effect on the galaxy overdensity field.
Following some early detections \citep{Bartelmann-Schneider-1994, Norman-Williams-2000},
the weak-lensing effect on clustering has been measured with high statistical significance ($8\sigma$) by cross correlating samples of distant quasars and background galaxies in the Sloan Digital Sky Survey \citep{Scrantonetal2005}.
Given this premise, we build a third set of mock light cones (labelled $\kappa$KD) in which we account for the redshift-space distortions due to both the peculiar velocities and weak lensing 
assuming that the convergence is the only source of magnification, i.e. ${\cal M}= 1+2\kappa$. To account for both magnification bias and the volume corrections due
to lensing, we simply weigh the N-body particles proportionally to $[{\cal M}(\hat{\bf n}_{\rm s},z)]^{{\cal Q}(z)-1}$ (instead of the standard $[{\cal M}(\hat{\bf n}_{\rm s},z)]^{{\cal Q}(z)}$) and use the same shifts as in the KD mocks.\footnote{
The additional ${\cal M}^{-1}$ results from a volume distortion due to lensing.
To test the consistency of our code, we evaluate this effect in two ways.
In the GR mocks the change of volume is realized by the particle shift.
Alternatively, we weigh particles proportionally to ${\cal M}^{-1}$ in the KD mocks.
We find the same result.}
The resulting spectra (yellow lines in Fig.~\ref{pic_angular_power}  which are barely distinguishable by eye from the blue ones) provide
an excellent fit to the $C_l^{(ij)}$ derived from the GR mocks. 
This suggests that the measurable differences between the full signal and the KD model are due to gravitational-lensing convergence.

In order to make a quantitative analysis and investigate whether other (more subtle and interesting) light-cone effects (e.g. Doppler lensing and potential terms) might be detectable with a Euclid-like survey, we once again resort to statistics. 
Fig.~\ref{pic_deltachi2_euclid_kkd} shows that,
in almost all mock GR realizations, the null hypothesis that the data are generated under the $\kappa$KD model cannot be rejected to any meaningful confidence level. 
Although only few of the skies presents deviations larger than $3\sigma$, Fig.~\ref{pic_deltachi2_euclid_kkd} indicates that the GR mocks contain an additional signal (most likely due to Doppler lensing) which is however comparable than the noise.
We thus conclude that no additional sources of redshift-space distortions beyond the Kaiser effect, Doppler contributions and magnification bias can be detected from the angular clustering of all galaxies with spectroscopic redshifts in a Euclid-like survey divided in four equally-populated redshift bins.
This, of course, does not prevent the development of dedicated probes to isolate additional contributions, in particular combining photometric and spectroscopic data to 
define multiple tracers of the large-scale structure along the lines of the forthcoming discussion in Section \ref{sect_lowred}.

\begin{figure}
  \centering
  \includegraphics[width=0.465\textwidth,bb=0 0 604 500,keepaspectratio=true]{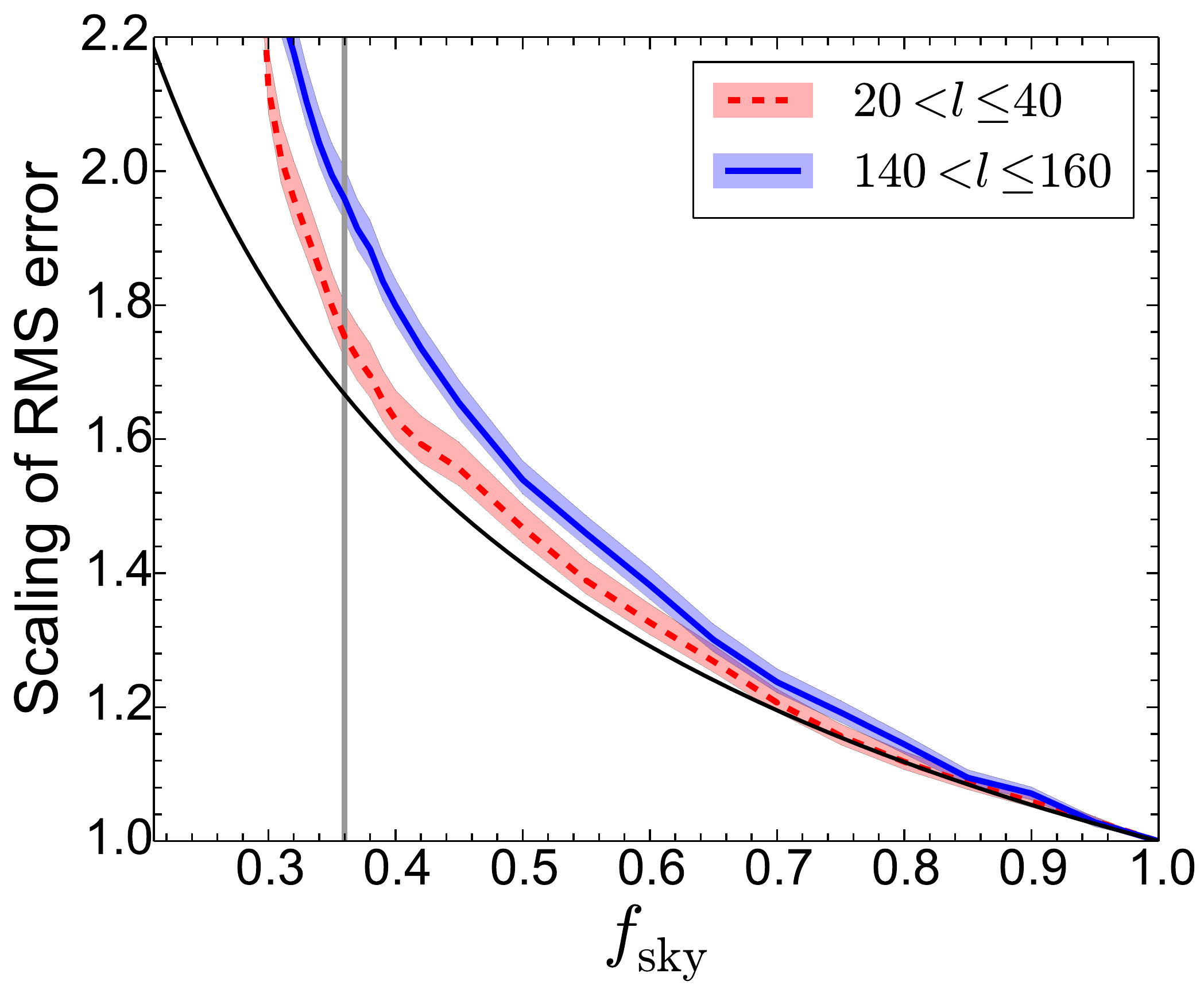}
  \caption{The average RMS statistical error of $\hat{C}_l^{(ij)}$ over the 165 Euclid mocks and 10 angular power spectra is plotted as a function of the covered sky fraction
  (all curves are normalized to unity for $f_{\rm sky}=1$). 
  The thick dashed and solid lines refer to two multipole bins centred at $l=30$ and $l=150$, respectively.
  The shaded areas highlight the corresponding uncertainty obtained bootstrapping the mock light cones.
  The black curve shows the scaling $\propto f_{\rm sky}^{-1/2}$ expected in the Gaussian approximation.
  The vertical grey line indicates the planned sky coverage of the Euclid redshift survey.}
  \label{pic_euclid_error}
\end{figure}

\subsection{Estimating covariances with LIGER}
Mock catalogues provide a direct way to estimate the covariance matrix of observables and test the range of validity of idealised theoretical models for the statistical errors and their
correlations.
The covariance matrix of the power spectrum on large scales is often described in terms of the Gaussian approximation,
\begin{eqnarray}\label{eq_covgaus}
 \Sigma_{lm}^{(ik)(jh)}=\delta_{lm}\,\frac{C_{\rm l}^{(ik)}C_{\rm l}^{(jh)}+C_{l}^{(ih)}C_{l}^{(jk)}}   {(2l+1)f_{\rm sky}}\;,
\end{eqnarray}
where $\delta_{lm}$ denotes the Kronecker symbol and we have restored the superscripts for the redshift bins, for clarity.
As a byproduct of the study presented in the previous section, we use the Euclid mock catalogues to assess the accuracy of the Gaussian approximation.  
Overall, our numerical estimates are compatible with the diagonal structure of $ \Sigma_{lm}^{(ik)(jh)}$.  
The diagonal elements obtained from the full-sky mocks are in excellent agreement with equation (\ref{eq_covgaus}).
However, we find that the expected scaling with respect to $f_{\rm sky}$ holds true only for low multipoles.
Fig.~\ref{pic_euclid_error} shows how the average standard deviation of $C_l^{(ij)}$ in the mocks varies as a function of 
$f_{\rm sky}$. 
The thick blue line corresponds to the multipole interval $20< l \leq40$, while the dashed line represents the bin $140<l\leq160$.
Both curves have been normalised by the corresponding standard deviations measured in the full-sky mocks.
The hatched area indicates the error on the ratio estimated bootstrapping the realisations.
The black line highlights the theoretical scaling proportional to $f_{\rm sky}^{-1/2}$.
Note that the statistical error for $C_l^{(ij)}$ at $l\sim 30$ follows this curve for $f_{\rm sky}>0.35$ but rapidly departs from it for smaller sky fractions. Basically,
the data cannot optimally constrain the large-scale power when the footprint of the survey covers too small a fraction of the sky.
Higher multipoles deviate from the ideal relation for even larger values of $f_{\rm sky}$.
For an Euclid-like survey, the statistical error on $C_l^{(ij)}$ at $l\sim 150$ is on average 17 per cent larger than expected using the $f_{\rm sky}^{-1/2}$ scaling. 
All this exemplifies the usefulness of LIGER (and mock catalogues in general) to estimate the size of measurement errors in clustering statistics
and warns against using simple approximations outside the range within which they have been accurately tested.

\begin{figure*}
 \centering
 \hfill
 \includegraphics[width=0.46\textwidth,bb=0 0 624 499,keepaspectratio=true]{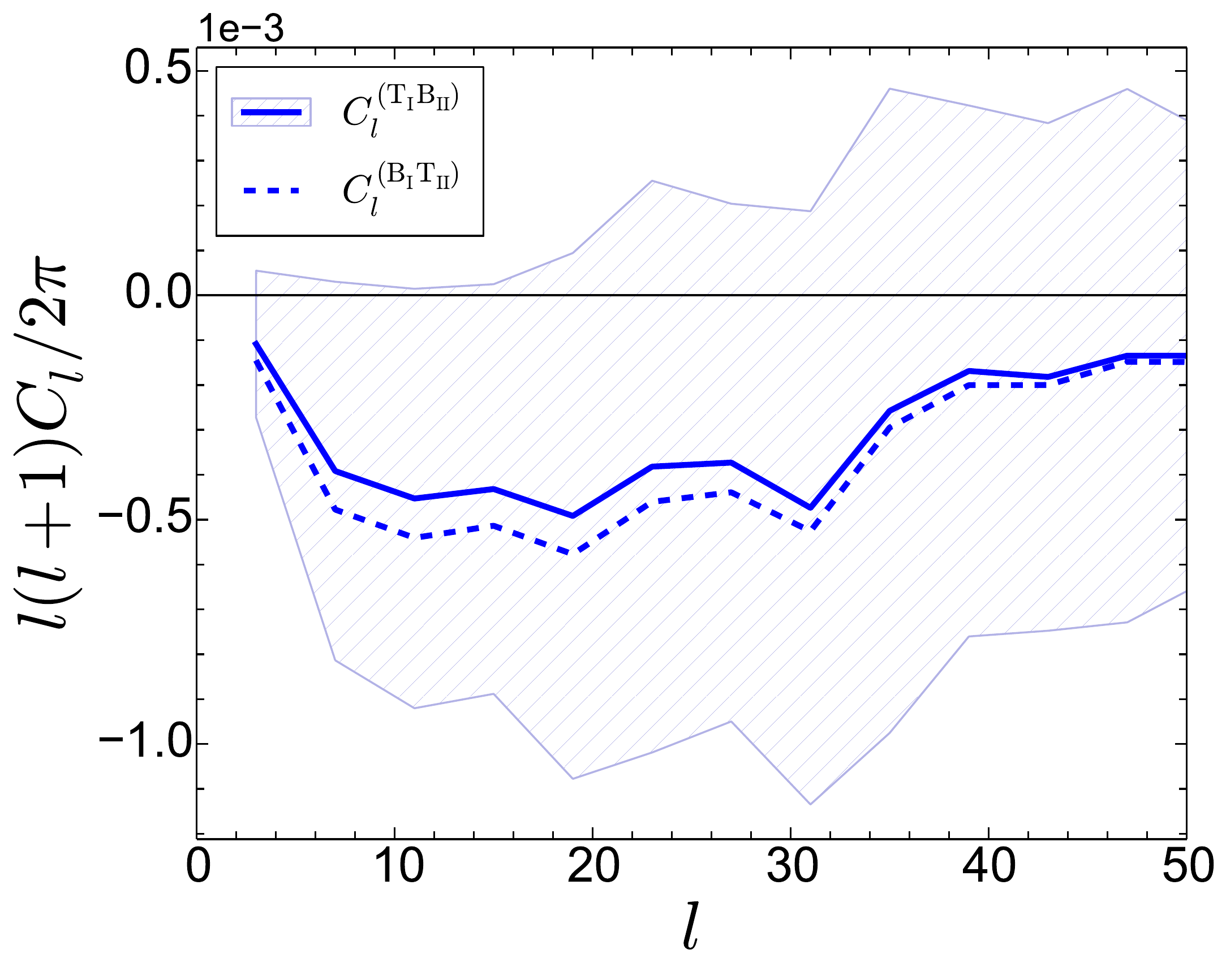}
 \hfill
 \includegraphics[width=0.46\textwidth,bb=0 0 624 499,keepaspectratio=true]{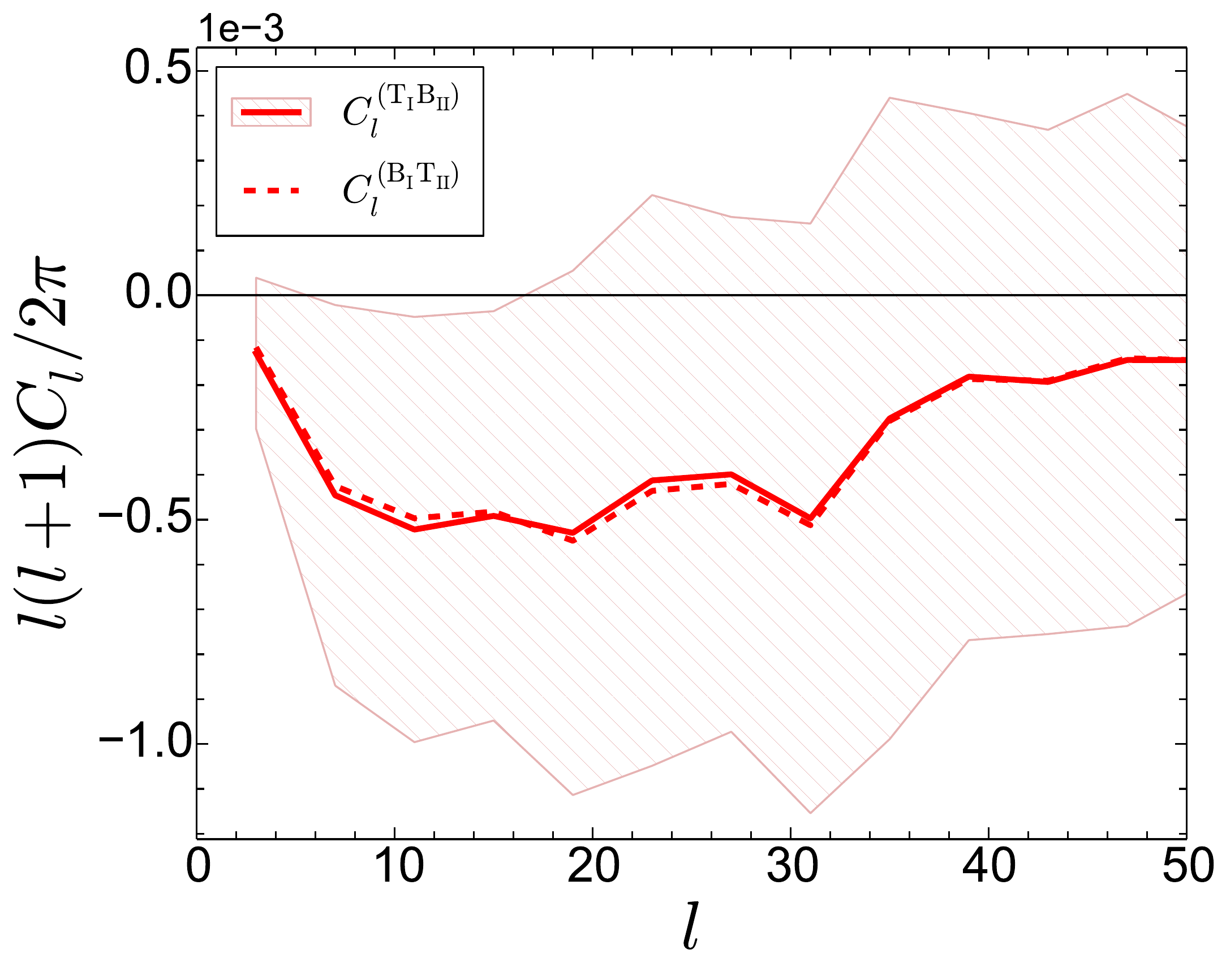}
 \hfill
 \caption{Forecasts for the billion-galaxy survey that will be conducted with the SKA2. We consider two galaxy populations
(the total sample (T) characterised by observed flux above $23\,\mu{\rm Jy}$ and the bright sub-sample (B) corresponding to a flux limit of $60\,\mu{\rm Jy}$)
sampled within two narrow and consecutive redshift bins (I: $0.15<z<0.2$ and II: $0.2<z<0.25$).
The average cross-spectra $\hat{C}_l^{\rm (T_I\, B_{II})}$ (solid thick line) and $\hat{C}_l^{\rm (B_I\, T_{II})}$ (dashed thick line) extracted from the full GR mocks are shown
in the left-hand panel. The hatched area denotes the standard deviation of the noise for $\hat{C}_l^{\rm (T_I\, B_{II})}$.
The corresponding quantities computed from the DS mock catalogues are displayed in the right-hand panel.}
 \label{pic_angular_power_run2}
\end{figure*}

\subsection{Detectability of Doppler terms at low redshift}\label{sect_lowred}
The question we want to address in this section is whether forthcoming probes of the large-scale structure will be able to provide evidence for additional sources
of redshift-space distortions beyond the classic Kaiser effect and magnification bias.
Although the term proportional to $\alpha$  in equation~(\ref{hamilton}) has been almost invariably neglected in past studies of galaxy clustering,
it might become detectable using data from future wide-angle surveys.  
Recent work based on analytical calculations has concluded that the Doppler terms should leave a measurable imprint on the
cross correlations between two galaxy populations \citep{McDonald2009, Yoo:2012se, Bonvinetal2014} and, possibly, also on the angular clustering of a single population \citep{Raccanelli:2016avd}.
Here we re-examine this case using LIGER to build mock catalogues for the SKA2 survey.

\subsubsection{Simulating a galaxy redshift survey with the SKA2}\label{sec_ska2}
The SKA is an unprecedentedly large and powerful array of radio telescopes that will be built in Australia and South Africa by an international collaboration.
The construction will be split into two phases: 10 per cent of the collective area should be in place by 2023 (SKA1) while the full array should follow by 2030 
(SKA2). 
The `billion galaxy survey' conducted with the SKA2 will measure $\sim 10^9$ individual galaxy redshifts 
over 30,000 deg$^2$ using the 21-cm line emission from neutral atomic hydrogen \citep{Maartens:2015SKA}.
In many senses, this will be the ultimate ground-based redshift survey for cosmology.

We build 125 mock light-cones with the expected characteristics of the billion-galaxy survey. 
Proceeding as described in Sections \ref{sect_galaxy_dist} and \ref{Nbodysims},
we populate our N-body simulations with galaxies making sure to reproduce 
the forecasts for the galaxy number counts presented in \cite{Yahyaetal2015}.
At low redshifts ($z\lesssim0.2$) and for flux\footnote{Although ${\rm Jy}$ is a unit of flux density for simplicity we refer to it as a flux.} limits below $10\,\mu{\rm Jy}$, their predictions for $\bar{n}_{\rm g}$ depend very little on the limiting flux of the survey 
(implying that $Q\simeq 0$). 
This might possibly reflect a shortcoming of their fitting formula (which covers a broad redshift range), an imperfection in the HI modeling, 
or the finite mass resolution of the N-body simulations used to make the forecasts (i.e. fainter galaxies might reside within unresolved dark-matter haloes).  
Indeed, \cite{Yahyaetal2015} note that observed HI mass functions at low redshifts contain more low-mass objects than found in the forecasts for SKA2. 
To play safe, we avoid this region of parameter space.
We thus use a conservative flux limit of $23\,\mu{\rm Jy}$ to define our main sample
\citep[][use $5.3\,\mu{\rm Jy}$ for their realistic forecasts and $23\,\mu{\rm Jy}$ for the pessimistic ones]{Yahyaetal2015} and
also consider a second galaxy population (the bright sample) with fluxes above $60\,\mu{\rm Jy}$.
We derive the corresponding values for $Q(z)$ and ${\cal E}(z)$ after fitting the cumulative number density presented in \citet{Yahyaetal2015} with a third-order polynomial.
This is the same approach followed by \citet{Cameraetal2015} and yields consistent results with their revised analysis\footnote{In a flux limited survey, $Q$ cannot assume negative values as it would imply that less galaxies are detected when a fainter limiting flux is considered 
(the presence of an additional bright cut or another selection criterion are necessary to drive $Q<0$).
Due to an unfortunate mishap, the original fitting functions presented in  \citet{Cameraetal2015} yield $Q<0$ for faint galaxies at low redshifts.
Although this mistake does not affect their conclusions, the negative magnification biases 
have been used by many authors to make forecasts for the SKA2 survey} \citep{Cameraetal2017}.
The outcome of our calculations is shown in Fig. \ref{Fig:surveydata} 
together with the functions $b(z)$ which we take directly from \citet{Yahyaetal2015}.

To test the impact of the Doppler terms on the clustering of SKA2 galaxies, we build and contrast two sets of mock catalogues. 
The first includes relativistic effects while the second drops the Doppler terms that are proportional to ${\cal E}$ and $Q$. 
We omit Doppler magnification by simply ignoring the velocity-dependent terms in equation~(\ref{eq_magni}). 
Further, we neglect the term proportional to ${\cal E}$ in equation~(\ref{emphdeltag}) so that 
the weight $w_{\rm g}(\bar{z})$ in the second line of equation~(\ref{eq_numberdens}) is replaced with $w_{\rm g}(z)$.
We use the label DS (Doppler suppressed) to indicate the light cones constructed in this way, since it is impossible to isolate the remaining Doppler effects.

\subsubsection{Cross spectra and results}
General relativistic corrections alter galaxy clustering on large scales with respect to the predictions of the `standard model' including
a linear bias and the Kaiser distortions.
In particular, they break the symmetry of two-point statistics under the exchange of particles in the pairs.
In the distant-observer approximation,
the relativistic effects generate odd multipoles in the redshift-space cross-correlation function between two galaxy populations or, equivalently, an imaginary part in the cross spectrum \citep{McDonald2009, Yoo:2012se,  Croft2013, Bonvinetal2014, Bonvin2014, Raccanellietal2014}.
In terms of the comoving wavenumber of the perturbations $k$,
relativistic corrections to the cross spectra due to Doppler effect and gravitational redshifts
are suppressed by a factor ${\cal H}/k$ with respect to the leading standard-model terms  \citep{McDonald2009}.
Additional corrections (due to the gravitational potential, the integrated Sachs-Wolfe effect and the Shapiro time delay) are instead suppressed by a factor $({\cal H}/k)^2$.
Measurements of galaxy clustering on scales comparable with the Hubble radius are therefore necessary to detect them.
There is a complication, however. In the standard model, 
the time evolution of the galaxy populations and wide-angle effects due to the fact we observe on our past light-cone give rise to several anti-symmetric terms with similar amplitudes to the relativistic corrections \citep{Bonvinetal2014}. Moreover, dust extinction can further introduce spurious anti-symmetric terms in the galaxy correlation functions \citep{Fang-Hui-etal-2011}. 
Finally, from the observational point of view, it is challenging to keep the photometry stable and measure weak clustering signals on very large scales.
Therefore, it is still an open question whether the Doppler contribution can be seen.
In the rest of this section, we will employ our SKA2 mock light cones to address this issue. In doing this, we will neglect systematic effects due to observational limitations and dust
and focus on the feasibility of the experiment from a theoretical point of view.

 \begin{figure}
 \centering
 \includegraphics [width=0.5\textwidth,bb=0 0 630 499,keepaspectratio=true]{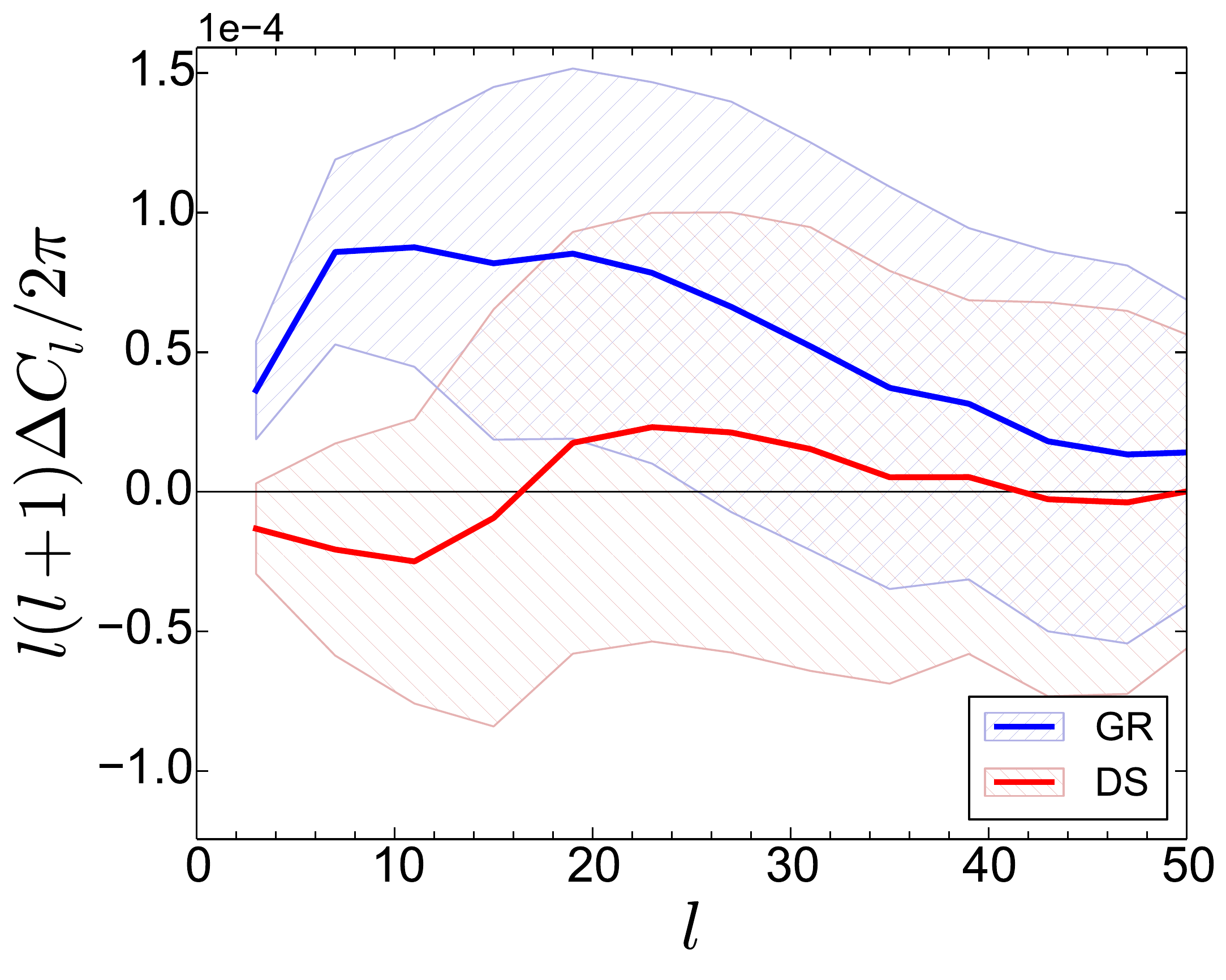}
 \caption{The observable $\Delta \hat{C}_l$ averaged over the GR (blue line) and DS (red line) mock catalogues. 
  Hatched areas indicate the standard deviation of the estimates over the realisations.}
 \label{pic_angular_power_difference}
\end{figure}

In order to maximise the chances for a successful outcome, it is important to carefully configure the test we want to perform.
There are a few facts to take into consideration: (i) linear peculiar velocities grow bigger at lower redshifts;
(ii) $\gamma_0$ in $\delta_{\rm g,v}$ is divided by ${\cal H}\chi_{\rm s}$; 
(iii) the galaxy number density of the samples rapidly decreases for $z\gtrsim0.2$ thus producing a large evolutionary bias but also increasing noise; 
(iv) we need to cover enough comoving volume to reduce sample variance.
Given all this, we end up  
considering the interval $0.15<z<0.25$ which we further divide into the bins
I: $0.15<z<0.2$ and II: $0.2<z<0.25$.
We also make use of the two galaxy populations introduced in Section~\ref{sec_ska2}.

\begin{figure*}
 \centering
 \includegraphics[width=0.945\textwidth,bb=0 0 1296 501,keepaspectratio=true]{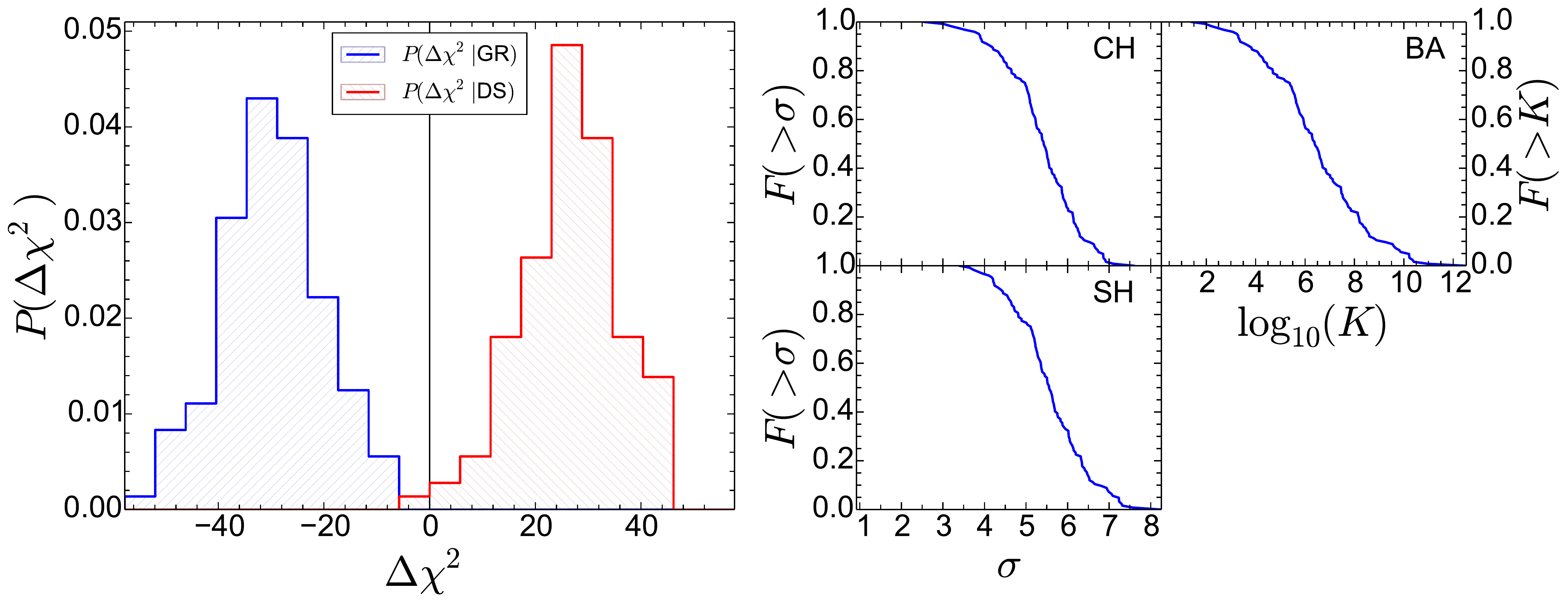}
 \caption{As in Fig.~\ref{pic_deltachi2_euclid} but for the 125 DS and GR mock catalogues for the billion-galaxy survey that will be conducted with the SKA2. 
 We fit the first 25 multipoles of $\Delta\hat{C}_l$ defined in equation~(\ref{eq_Cldiff}).
 The fact that the histograms do not significantly overlap demonstrates that the SKA2 will be able to detect non-standard Doppler terms in the galaxy-clustering signal.}
 \label{pic_deltachi2_ska2}
\end{figure*}

In order to access the scales that are sensitive to the Doppler contribution, we focus on galaxy pairs with very wide angular separations.
We thus compute the cross angular power spectrum, $\hat{C}_l^{\rm (T_I\, B_{II})}$, between the total galaxy sample within the lower redshift bin and the bright subsample within 
the highest redshift bin.
Similarly, we measure $\hat{C}_l^{\rm (B_I\, T_{II})}$ by considering the bright galaxies in bin I and the full population in bin II.
Finally, we consider the difference between the cross spectra:
\begin{eqnarray}\label{eq_Cldiff}
 \Delta\hat{C}_l=\hat{C}_l^{\rm (T_I\,B_{II})}-\hat{C}_l^{\rm (B_I\,T_{II})}\;.
\end{eqnarray}
Average results over the SKA2 mock light cones are shown in Fig.~\ref{pic_angular_power_run2}.
The left-hand panel displays $\hat{C}_l^{\rm (T_I\, B_{II})}$ (solid thick line) and $\hat{C}_l^{\rm (B_I\, T_{II})}$ (dashed thick line) both sampled in bandpowers with $\Delta l=4$.
The same quantities derived from the DS mock catalogues are shown in the right-hand panel.
The additional Doppler terms included in the GR light cones clearly enhance the difference between the cross spectra at low $l$. They boost  $\hat{C}_l^{\rm (B_I\, T_{II})}$ (i.e. make it more negative)
and suppress $\hat{C}_l^{\rm (T_I\, B_{II})}$.
This can be explained as follows.
The leading Doppler contribution to the cross spectra 
on large scales originates from correlating the term $b \delta_{\rm sim}-{\cal H}^{-1}\partial n_{\rm s}^iv_i/\partial\chi_{\rm s}$ with
the Doppler term in equation (\ref{hamilton}). Therefore, schematically, in terms of single Fourier modes,
\begin{eqnarray}
\Delta \hat{C}_l\!\!\!\!\!&\propto&\!\!\!\!\!\left[(b_{\rm T_I}+\mu_{\rm I}^2f_{\rm I}) \alpha_{\rm B_{II}}-(b_{\rm B_I}+\mu_{\rm I}^2f_{\rm I})\alpha_{\rm T_{II}}\right]\frac{\delta_{\rm I}\,v_{\rm II}}{\cH_{\rm II}\chi_{\rm II}}\nonumber\\
&+&\!\!\!\!\!\left[(b_{\rm B_{II}}+\mu_{\rm II}^2f_{\rm II})\alpha_{\rm T_{I}}-(b_{\rm T_{II}}+\mu_{\rm II}^2f_{\rm II})\alpha_{\rm B_I}\right]\frac{\delta_{\rm II}\,v_{\rm I}}{\cH_{\rm I}\chi_{\rm I}} \;,
\end{eqnarray}
where $f=\ud \ln  D/\ud \ln a$ (with $D$ the linear growth factor of matter perturbations) describes the evolution of the velocity field, while $\mu_{\rm I}$ and $\mu_{\rm II}$ are the cosines between the wavenumber and the two lines of sight in the wide-angle configuration.
A peculiarity of our samples is that $b_{\rm B}(z)\simeq b_{\rm T}(z)=b(z)$ to very good accuracy \citep[this is different from][which relies on
different linear bias parameters and use three dimensional correlation functions]{Bonvinetal2014} so that 
\begin{eqnarray}
\Delta \hat{C}_l\!\!\!\!\!&\propto&\!\!\!\!\!\left(\alpha_{\rm B_{II}}-\alpha_{\rm T_{II}}\right)\left(b_{\rm I}+\mu_{\rm I}^2f_{\rm I} \right)\frac{\delta_{\rm I}\,v_{\rm II}}{\cH_{\rm II}\chi_{\rm II}}\nonumber\\
&+&\!\!\!\!\! \left(\alpha_{\rm T_{I}}-\alpha_{\rm B_I}\right)\left(b_{\rm II}+\mu_{\rm II}^2f_{\rm II}\right)\frac{\delta_{\rm II}\,v_{\rm I}}{\cH_{\rm I}\chi_{\rm I}}\;.
 \end{eqnarray}
Now,
the DS mocks only include the geometric distortions and those generated by cosmic acceleration, i.e. $\alpha_{\rm DS}=2+[1-(3/2)\Omega_{\rm m}(z)] \cH\chi_{\rm s}$ which
is independent of the galaxy sample and only depends on the underlying cosmological model. Thus, $\Delta \hat{C}_l \simeq 0$ in the DS case.
On the other hand, for the GR mocks we get
 $\alpha_{\rm B_{I}}=-1.49$, $\alpha_{\rm B_{II}}=-2.42$, $\alpha_{\rm T_{I}}=0.35$ and $\alpha_{\rm T_{II}}=-0.38$ which combine to produce a positive signal.
In this case ${\cal Q}$ and ${\cal E}$ give roughly equal contributions.

The shaded areas in Fig.~\ref{pic_angular_power_run2} indicate the standard deviation for $\hat{C}_l^{\rm (T_{\rm I}\, B_{\rm II})}$ (the scatter for $\hat{C}_l^{\rm (B_{\rm I}\, T_{\rm II})}$ is of comparable size).
It is evident that the cross spectra extracted from the galaxy survey will be very noisy.
In Fig.~\ref{pic_angular_power_difference} we show the mean and scatter of $\Delta\hat{C}_l$ for the 
full GR mocks (blue) and the DS ones (red). 
While the relative error on the single cross-spectra is very large, $\Delta\hat{C}_l$ can be measured with much better accuracy (especially for $l<25$).
Since both galaxy populations trace the same large-scale structure, most of the noise in the cross-spectra is correlated and thus does not appear in the difference.
This exemplifies the advantage of using a multi-tracer approach \citep{McDonald-Seljak-2009}.

We are now ready to investigate whether Doppler effects will be measurable with the SKA2. 
We first measure $\Delta\hat{C}_l$ from the 125 GR mocks and then fit the multipoles in the range $1\le l\le25$ using both the GR and the DS models (we recall that the models are obtained averaging the signal over all the mock light cones).
The left-hand side histogram in the left panel of Fig.~\ref{pic_deltachi2_ska2} shows the distribution of the corresponding value for $\Delta \chi^2=\chi^2_{\rm GR}-\chi^2_{\rm DS}$. 
In parallel, we fit the $\Delta\hat{C}_l$ measurements extracted from the DS mocks and plot the corresponding histogram of $\Delta \chi^2$ on the right-hand side of the figure.
The fact that the two histograms are well separated demonstrates that the SKA2 data should contain enough information to distinguish between the two models.
To better quantify this, for each mock light cone we compute the statistical significance with which we can reject the null hypothesis that the GR data are generated by the DS model.
The resulting cumulative distribution is plotted in the right panel of Fig.~\ref{pic_deltachi2_ska2} using three different statistical tests (see Section~\ref{sect_significance}).
In the vast majority of the mock catalogues, the DS model is ruled out at a confidence ranging between 4 and 7$\sigma$ (or with a Bayes factor above 1000 corresponding
to decisive evidence).
We conclude that the SKA2 should be able to detect the non-standard Doppler contribution to galaxy clustering.

Our results heavily rely upon the multi-tracer technique for the suppression of the statistical noise. 
After repeating the analysis with a single tracer (using cross correlations between adjacent and narrow redshift bins), we find that only extreme values for the magnification and evolutionary biases  (e.g. $Q \gtrsim 10$ and $|{\cal E}|\gtrsim20$) could lead to a statistically significant detection. 
 
\section{Summary \& conclusions} 
There are multiple reasons for which galaxy clustering requires a proper general relativistic description. 
 (i) We observe events lying on our past light cone.  
 (ii) The propagation of light is affected by the presence of inhomogeneities in the matter distribution. 
 (iii) In consequence, galaxy observables (i.e. redshift, flux in some waveband and angular position on the sky) are influenced by the large-scale structure intervening between the source and the observer. 
 However, when we interpret observations we use an unperturbed FRW model to translate redshifts and fluxes into distances and absolute luminosities. 
 This leads to redshift-space distortions, i.e. the reconstructed galaxy density does not coincide with the actual one. 
 The most important source of the discrepancy is the correction due to the peculiar velocity gradient \citep{Kaiser:1987qv} 
 but it is long known that there are additional contributions and that they might become significant at large angular separations. 
 Although no consensus has been reached yet, some recent studies based on analytical calculations and on the Fisher information matrix  
 have concluded that signatures of these additional corrections should be detectable with the next generation of wide-angle surveys. 
  
 This paper describes the LIGER method, 
 a numerical technique to build mock galaxy catalogues including all general relativistic corrections at linear order in the cosmological perturbations. 
 LIGER post processes the output of a Newtonian simulation and combines its snapshots at constant background time 
 to build the galaxy distribution in comoving redshift space. 
 The LIGER method is composed of three steps: 
 (i) we shift the position of the simulated galaxies from real to redshift space; 
 (ii) we evaluate the corresponding magnification due to gravitational lensing; 
 (iii) we find the intersection of the modified world lines of the galaxies with the unperturbed backward light cone of the observer in redshift space. 
Steps (i) and (ii) include both local corrections and terms that have been integrated along the line of sight using the Born approximation.  
Note that standard mock galaxy catalogues generally include only the Kaiser effect for step (i) and do not account for lensing 
 \citep[with the exception of][]{Guimaraesetal2005} although magnification maps are routinely built for  weak-lensing studies  
  \citep[e.g.][]{Wambsganss1998,Jainetal2000,WhiteHu2000,ValeWhite2003,Fosalbaetal2008, Fosalbaetal2015,Hilbertetal2009,Kiesslingetal2011}. 
 
 LIGER is characterized by the following features.   
 (i) It can be applied to the output of any Newtonian simulation (N-body or hydrodynamic) independent of the 
 code with which it has been run. (ii) It is fast to execute so that it is computationally feasible to produce very large numbers of mock catalogues for a given 
 survey. 
 (iii) A variant of the standard implementation has been specially designed to work with simulations that cover very large comoving volumes but 
 do not resolve single galaxies. In this case, the galaxy density field is obtained by biasing the dark-matter distribution. 
  
 The main applications of LIGER are foreseen for forthcoming wide-angle spectroscopic surveys like  
 DESI\footnote{http://desi.lbl.gov},  
 Euclid\footnote{http://www.euclid-ec.org},  
 HETDEX\footnote{http://hetdex.org},  
 SKAII\footnote{https://www.skatelescope.org}, 
 SPHEREx\footnote{http://spherex.caltech.edu}, 
 SuMiRe\footnote{http://sumire.ipmu.jp/en/} 
 as well as photometric surveys like J-PAS\footnote{http://j-pas.org} and LSST\footnote{https://www.lsst.org}. 
Mock catalogues will be used as forecasting tools and to measure biases and covariance matrices of estimators for several statistics of the large-scale structure.  
As a proof of concept, we quantify the impact of magnification bias in the angular clustering of galaxies for a Euclid-like survey. 
Our results show that lensing convergence generates a non-negligible correction for the angular power spectra of galaxies in broad redshift bins and 
dominates cross-spectra between galaxies at widely separated redshifts
 \citep[similar conclusions have been reached by][using analytical calculations]{DiDioetal2014, Montanari:2015rga}. 
 The convergence signal can be detected at 8$\sigma$ significance and
 this provides the intriguing possibility to measure the lensing potential from the cross spectra \citep{Montanari:2015rga}. 
 
Additional redshift-space distortions on top of the standard Kaiser correction and the weak-lensing convergence   
are generally small. Therefore, customised techniques need be developed in order to measure their signatures.  
It is foreseeable that LIGER-based mocks will be key to optimising the design of these probes. 
In anticipation of these future applications, we have investigated the detectability of several additional Doppler terms using the specifics of  
the planned `billion galaxy survey' with the SKA2 telescope.  
Our results show that using two galaxy populations with different flux cuts (T: $f>23 \,\mu{\rm Jy}$ and B: $f>60 \,\mu{\rm Jy}$)  
and two consecutive redshift intervals (I: $0.15<z<0.2$ and II: $0.2<z<0.25$),  it will be possible to measure a significant Doppler-induced signal. 
The statistic we use is the difference of the angular cross power spectra $\Delta C_l=C_l^{\rm (T_I\,B_{II})}-C_l^{\rm (B_I\, T_{II})}$.  
For multipoles $l\le25$, this quantity is dominated by the contribution of the Doppler terms and
shows a strikingly reduced variance compared to each of the cross correlations due to the fact that  
both galaxy populations are biased tracers of the same underlying matter density. Based on our simulations, $\Delta C_l$ should be detectable with a  
signal-to-noise ratio of $\sim 5.5$. 
 
\section*{Acknowledgements} 
We thank Alvise Raccanelli and Stefano Camera for discussions.
We acknowledge financial support by the Deutsche Forschungsgemeinschaft through the Transregio 33 `The Dark Universe'.  
During the preparation of this work MB was also partially supported by the Bonn-Cologne Graduate School for Physics and Astronomy. 
 
\bibliographystyle{mn2e}  
 \bibliography{paper}{}   
 
\appendix 
\section{Likelihood-ratio test}\label{Appendix} 
Let ${\mathbf x}$ be a Gaussian data vector with mean ${\mathbf m}$ and covariance matrix $\mathmat{\Sigma}$. 
Let ${\mathbf M}$ be a perfect theoretical model for ${\mathbf m}$, i.e. ${\mathbf M}={\mathbf m}$. 
The likelihood of ${\mathbf M}$ given the data is proportional to the probability of observing the data under the hypothesis that ${\mathbf M}$ is true, i.e. 
${\cal L}({\mathbf M}|{\mathbf x})\propto \exp(-\chi^2/2)$ with $\chi^2=({\mathbf x}-{\mathbf M})^{\rm T}\cdot\mathmat{\Sigma}^{-1}\cdot({\mathbf x}-{\mathbf M})$. 
Let also consider an imperfect or incomplete model ${\mathbf N}$ such that ${\mathbf m}-{\mathbf N}={\mathbf B}$. The likelihood ratio between ${\mathbf M}$ and ${\mathbf N}$ is 
${\cal L}({\mathbf M}|{\mathbf x})/{\cal L}({\mathbf N}|{\mathbf x})=\exp(-\Delta \chi^2/2)$ with 
\begin{eqnarray} 
\Delta \chi^2\!\!\!\!\!&=&\!\!\!\!\!({\mathbf x}-{\mathbf N})^{\rm T}\cdot\mathmat{\Sigma}^{-1}\cdot({\mathbf x}-{\mathbf N})-({\mathbf x}-{\mathbf M})^{\rm T}\cdot\mathmat{\Sigma}^{-1}\cdot({\mathbf x}-{\mathbf M})\nonumber \\ 
&=&\!\!\!\!\! ({\mathbf B}+{\mathbf e})^{\rm T}\cdot\mathmat{\Sigma}^{-1}\cdot({\mathbf B}+{\mathbf e})-{\mathbf e}^{\rm T}\cdot\mathmat{\Sigma}^{-1}\cdot{\mathbf e}\nonumber \\ 
&=&\!\!\!\!\! {\mathbf B}^{\rm T}\cdot\mathmat{\Sigma}^{-1}\cdot{\mathbf B}+2\,{\mathbf e}^{\rm T}\cdot\mathmat{\Sigma}^{-1}\cdot{\mathbf B}\;, 
\end{eqnarray} 
where we have decomposed the data vector in the signal and noise components, ${\mathbf x}={\mathbf m}+{\mathbf e}={\mathbf M}+{\mathbf e}$, and used the symmetry 
of the covariance matrix. 
The fact that the difference in the log-likelihoods depends linearly on ${\mathbf e}$ implies that $\Delta \chi^2$ follows a Gaussian distribution 
over an ensemble of realisations of the data vector. Its expectation is 
\begin{equation} 
\mu=E[\Delta \chi^2]={\mathbf B}^{\rm T}\cdot\mathmat{\Sigma}^{-1}\cdot{\mathbf B}\;. 
\end{equation} 
Similarly, its variance is 
\begin{equation} 
E[(\Delta \chi^2-\mu)^2]=4\,\mu. 
\end{equation} 
In a classic likelihood-ratio test for a simple hypothesis, the probability distribution of $\Delta \chi^2$ obtained under a model 
(i.e. assuming that this model perfectly describes ${\mathbf m}$) 
is compared against the corresponding distribution obtained under an alternative model. 
If the covariance matrix of the data does not depend on the model, the two $\Delta \chi^2$ distributions differ only in the sign of their mean values (in fact 
models $M$ and $N$ are switched in the alternative hypothesis and the sign of $\Delta \chi^2$ is reversed).  The test rejects one of the models if the two distributions of $\Delta \chi^2$ are clearly separated with respect to their intrinsic dispersion. Since the distance between the averages is $2\mu$ and the RMS value of each distribution is $2\sqrt{\mu}$, 
the ratio $2\mu/(2\sqrt{\mu})=\sqrt{\mu}$ is commonly referred to as the signal-to-noise ratio $S/N$. If, however, the covariance matrix is model dependent, then the comparison should be done between two Gaussian distributions with 
different mean values $\mu_M={\mathbf B}^{\rm T}\cdot\mathmat{\Sigma}_M^{-1}\cdot{\mathbf B}$ and $\mu_N=-{\mathbf B}^{\rm T}\cdot\mathmat{\Sigma}_N^{-1}\cdot{\mathbf B}$ as well as 
variances $4\mu_M$ and $4\mu_N$, respectively.

\label{lastpage} 
 
\end{document}

%% file: adsmacro.tex
%
%
%
%


\def\refads@jnl#1{{\rm#1}}

\def\aj{\refads@jnl{AJ}}                   
\def\actaa{\refads@jnl{Acta Astron.}}      
\def\araa{\refads@jnl{ARA\&A}}             
\def\apj{\refads@jnl{ApJ}}                 
\def\apjl{\refads@jnl{ApJ}}                
\def\apjs{\refads@jnl{ApJS}}               
\def\ao{\refads@jnl{Appl.~Opt.}}           
\def\apss{\refads@jnl{Ap\&SS}}             
\def\aap{\refads@jnl{A\&A}}                
\def\aapr{\refads@jnl{A\&A~Rev.}}          
\def\aaps{\refads@jnl{A\&AS}}              
\def\azh{\refads@jnl{AZh}}                 
\def\baas{\refads@jnl{BAAS}}               
\def\bac{\refads@jnl{Bull. astr. Inst. Czechosl.}}
\def\caa{\refads@jnl{Chinese Astron. Astrophys.}}
\def\cjaa{\refads@jnl{Chinese J. Astron. Astrophys.}}
\def\icarus{\refads@jnl{Icarus}}           
\def\jcap{\refads@jnl{J. Cosmology Astropart. Phys.}}
\def\jrasc{\refads@jnl{JRASC}}             
\def\memras{\refads@jnl{MmRAS}}            
\def\mnras{\refads@jnl{MNRAS}}             
\def\na{\refads@jnl{New A}}                
\def\nar{\refads@jnl{New A Rev.}}          
\def\pra{\refads@jnl{Phys.~Rev.~A}}        
\def\prb{\refads@jnl{Phys.~Rev.~B}}        
\def\prc{\refads@jnl{Phys.~Rev.~C}}        
\def\prd{\refads@jnl{Phys.~Rev.~D}}        
\def\pre{\refads@jnl{Phys.~Rev.~E}}        
\def\prl{\refads@jnl{Phys.~Rev.~Lett.}}    
\def\pasa{\refads@jnl{PASA}}               
\def\pasp{\refads@jnl{PASP}}               
\def\pasj{\refads@jnl{PASJ}}               
\def\rmxaa{\refads@jnl{Rev. Mexicana Astron. Astrofis.}}%
\def\qjras{\refads@jnl{QJRAS}}             
\def\skytel{\refads@jnl{S\&T}}             
\def\solphys{\refads@jnl{Sol.~Phys.}}      
\def\sovast{\refads@jnl{Soviet~Ast.}}      
\def\ssr{\refads@jnl{Space~Sci.~Rev.}}     
\def\zap{\refads@jnl{ZAp}}                 
\def\nat{\refads@jnl{Nature}}              
\def\iaucirc{\refads@jnl{IAU~Circ.}}       
\def\aplett{\refads@jnl{Astrophys.~Lett.}} 
\def\apspr{\refads@jnl{Astrophys.~Space~Phys.~Res.}}
\def\bain{\refads@jnl{Bull.~Astron.~Inst.~Netherlands}} 
\def\fcp{\refads@jnl{Fund.~Cosmic~Phys.}}  
\def\gca{\refads@jnl{Geochim.~Cosmochim.~Acta}}   
\def\grl{\refads@jnl{Geophys.~Res.~Lett.}} 
\def\jcp{\refads@jnl{J.~Chem.~Phys.}}      
\def\jgr{\refads@jnl{J.~Geophys.~Res.}}    
\def\jqsrt{\refads@jnl{J.~Quant.~Spec.~Radiat.~Transf.}}
\def\memsai{\refads@jnl{Mem.~Soc.~Astron.~Italiana}}
\def\nphysa{\refads@jnl{Nucl.~Phys.~A}}   
\def\physrep{\refads@jnl{Phys.~Rep.}}   
\def\physscr{\refads@jnl{Phys.~Scr}}   
\def\planss{\refads@jnl{Planet.~Space~Sci.}}   
\def\procspie{\refads@jnl{Proc.~SPIE}}   

\let\astap=\aap
\let\apjlett=\apjl
\let\apjsupp=\apjs
\let\applopt=\ao